\documentclass[psfig]{JHEP}

\usepackage{amssymb,epsf}

\newcommand{\be}{\begin{equation}}
\newcommand{\ee}{\end{equation}}
\newcommand{\bea}{\begin{eqnarray}}
\newcommand{\eea}{\end{eqnarray}}

\newcommand{\ds}{\displaystyle}
\newcommand{\half}{{1\over2}}

\newcommand{\fa}{\forall\,}
\renewcommand{\a}{\"a}
\renewcommand{\o}{\"o}
\renewcommand{\u}{\"u}
\newcommand\C{\mathbb{C}}
\renewcommand\H{\mathbb{H}}
\newcommand\R{\mathbb{R}}
\newcommand\Z{\mathbb{Z}}
\newcommand\N{\mathbb{N}}
\newcommand\mod{\mathop{\mathrm{mod}\,}}

\newcommand{\crl}{fourteen }
\newcommand{\crit}{31 }
\newcommand{\qcr}{three }
\newcommand{\tcr}{ten }

\def\sqr#1#2{{\vcenter{\vbox{\hrule height.#2pt
            \hbox{\vrule width.#2pt height#1pt \kern#1pt
                  \vrule width.#2pt}\hrule height.#2pt}}}}
\def\square{
\mathop{\mathchoice{\sqr{13}{16}}{\sqr{12}{15}}{\sqr{8}{10}}{\sqr{4}{5}}}}
\newcommand\mbar[1]{\raisebox{0.2ex}
                   {$\stackrel{\mbox{\tiny(}-\mbox{\tiny)}}{#1}$}}
\newcommand\mpbar[1]{\raisebox{-0.3ex}
                    {$\stackrel{\mbox{\tiny(}-\mbox{\tiny)}}{#1}$}}
\newcommand\pmm{\raisebox{-0.7ex}{$\stackrel{\ds+}
                        {\mbox{\tiny(}\!-\!\mbox{\tiny)}}$}\,}
\newcommand\diag{\mathop{\mathrm{diag}}}
\newcounter{quadru}

\newenvironment{quadru}[1][]{
\refstepcounter{quadru}\addtocounter{equation}{-1}

\begin{eqnarray}}{\end{eqnarray}}
\newcounter{triple}

\newenvironment{triple}[1][]{
\refstepcounter{triple} \addtocounter{equation}{-1}

\begin{equation}\label{#1}
}{\end{equation}}
\newcommand\bt[1]{\begin{triple}[#1]}
\newcommand\et{\end{triple}}
\newcounter{double}

\newenvironment{double}[1][]{
\refstepcounter{double} \addtocounter{equation}{-1}

\begin{equation}\label{#1}
}{\end{equation}}
\newcommand\bd[1]{\begin{double}[#1]}
\newcommand\ed{\end{double}}
\newcommand\dlabel[1]{\refstepcounter{double}\addtocounter{equation}{-1}
\label{#1}}
\newenvironment{doubles}{
\addtocounter{equation}{-1}

\begin{eqnarray}
}{\end{eqnarray}}
\newcommand\bds{\begin{doubles}}
\newcommand\eds{\end{doubles}}
\newcounter{crline}

\newenvironment{crline}[1][]{
\refstepcounter{crline} \addtocounter{equation}{-1}

\begin{equation}\label{#1}
}{\end{equation}}
\newcommand\bl[1]{\begin{crline}[#1]}
\newcommand\el{\end{crline}}
\newcommand\llabel[1]{\refstepcounter{crline}\addtocounter{equation}{-1}
\label{#1}}
\newenvironment{crlines}{

\begin{eqnarray}
}{\end{eqnarray}}
\newcommand\bls{\begin{crlines}}
\newcommand\els{\end{crlines}}

\newcommand{\Gl}{\mathop{\rm Gl}}
\newcommand{\Sl}{\mathop{\rm Sl}}
\newcommand{\SU}{\mathop{\rm SU}}
\newcommand{\U}{\mathop{\rm {}U}}
\newcommand{\OO}{\mathop{\rm {}O}}
\newcommand{\uu}{\mathop{\rm {}u}}
\newcommand{\PSL}{\mathop{\rm PSL}}
\newcommand{\Skew}{\mathop{\rm Skew}}
\renewcommand{\Im}{\mathop{\rm Im}\nolimits}
\newcommand{\tr}{\mathop{\rm tr}\nolimits}

\title{Crystallographic orbifolds: towards a classification of unitary
conformal field theories with central charge $c = 2$}

\author{Sayipjamal Dulat and Katrin Wendland\\
Physikalisches Institut, Universit\a t Bonn\\
Nu\ss allee 12, D-53115 Bonn, Germany\\
E-mail: \email{sayip@th.physik.uni-bonn.de},
	\email{wendland@th.physik.uni-bonn.de}}

\abstract{We study the moduli space $\mathcal{C}^2$ of unitary
two-dimensional conformal field theories with central charge $c=2$. We
construct all the 28 nonexceptional nonisolated irreducible components
of $\mathcal{C}^2$ that may be obtained by an orbifold procedure from
toroidal theories. The parameter spaces and partition functions are
calculated explicitly, and all multicritical points and lines are
determined. We show that all but four of the 28 irreducible components
of $\mathcal{C}^2$ corresponding to nonexceptional orbifolds are
directly or indirectly connected to the moduli space of toroidal
theories in $\mathcal{C}^2$.  We relate our results to those by Dixon,
Ginsparg, Harvey on the classification of $c=3/2$ superconformal field
theories and thereby give geometric interpretations to all nonisolated
orbifolds discussed there.}

\preprint{hep-th/0002227}

\begin{document}

\section{Introduction}

\looseness=1 In this paper we study the moduli space $\mathcal{C}^2$ of unitary
two-dimensional conformal field theories with central charge $c=2$.
The component $\mathcal{T}^2$ of the moduli space corresponding to
compactification on a two-dimensional torus is well
understood~\cite{cent,na}.  One can conjecture that every theory in
$\mathcal{C}^2$ either corresponds to compactification on a torus or
on an orbifold thereof.  It was stated in~\cite{dijk} that it is not
difficult to classify all possible types of $c=2$ (symmetric) orbifold
models which can be obtained by modding out an automorphism group of a
theory in $\mathcal{T}^2$.  However, to our knowledge this analysis
has not been carried out explicitly up to now.

The paper is organized as follows: in section~\ref{tori} we briefly
review the features of $\mathcal{T}^2$ relevant to our studies.
Moreover, we argue that apart from some exceptional cases any
nonisolated component of $\mathcal{C}^2$ which can be constructed by
applying an orbifold procedure to a subspace of the Teichm\u ller
space of $\mathcal{T}^2$ can be obtained by modding out an
automorphism group of a two-dimensional torus.  This means that to
find all such nonisolated components we can use the standard
classification of crystallographic groups in two dimensions, which is
discussed in section~\ref{symm}.  Section~\ref{part} contains a case
by case study of all the 28 irreducible components of $\mathcal{C}^2$
obtained from $\mathcal{T}^2$ by modding out crystallographic groups.
All consistent choices of the B-field on the original toroidal theory
and the effect of discrete torsion are discussed, which also leads to
some insight into the role of the B-field in a conformal field theory.
We explicitly calculate the corresponding partition functions and
determine the parameter space for each component.  In
section~\ref{meet} we make use of results of
B.~Rostand's~\cite{ro90,ro91} to determine all intersections of the
irreducible components of the moduli space obtained in
section~\ref{part}, i.e.\ singular or multicritical lines and points in
$\mathcal{C}^2$.  This also sheds some light on the effect of discrete
torsion.  We find a whole wealth of \crl multicritical lines and \crit
multicritical points for the crystallographic components, among them
\qcr quadrucritical and \tcr tricritical points.  In particular, we
show that all but four of the components of $\mathcal{C}^2$
constructed by crystallographic orbifolds are directly or indirectly
connected to $\mathcal{T}^2$.  The moduli space exhibits a complicated
graph like structure with many loops.  In section~\ref{product} we
discuss theories obtained as tensor products of known models with
central charge $c<2$. We relate our results to those on $c=3/2$
superconformal field theories~\cite{dix} and are able to interprete
all the orbifolds discussed there in terms of crystallographic
orbifolds.

Unitary two-dimensional quantum field theories can be described as
minkowskian theories on the circle or equivalently as euclidean
theories on the torus with parameter $\sigma$ in the upper half plane.
The world sheet coordinates are called $\xi_0, \xi_1$, and we
frequently use $z=e^{\xi_0+\sigma\xi_1}$, $z\in Z$ to parametrize the
worldsheet on an annulus $Z\subset\C^\ast$.

\section{The moduli space $\mathcal{T}^2$ of toroidal theories}\label{tori}

Let us briefly recall the structure of the moduli space
$\mathcal{T}^2$ of theories corresponding to toroidal compactification
in two dimensions (see also~\cite{dijk}).  Consider a torus
$\mathbb{T}^2=\R^2/\Lambda$, where $\Lambda\subset\R^2$ is a nondegenerate
lattice with generators $\lambda_1, \lambda_2\in\Lambda$.  The
nonlinear $\sigma$-model on $\mathbb{T}^2$ describes two real massless scalar
fields $\Phi^\mu: Z\rightarrow \mathbb{T}^2$, $\mu\in\{1,2\}$, governed by the
action
\be\label{boac}
S =\frac{1}{2\pi}\int_Z d^2z\;(G_{\mu\nu} + B_{\mu\nu}) 
   \partial\Phi^{\mu}(z,\bar z)\bar{\partial}\Phi^{\nu}(z,\bar z) \,,
\ee
where we have set $\alpha^\prime=1$ by choosing a unit of length.  The
constant symmetric tensor
$G_{\mu\nu}=\langle\lambda_\mu,\lambda_\nu\rangle$ defines the metric
on $\mathbb{T}^2$ and the antisymmetric tensor $B_{\mu\nu}=-B_{\nu\mu}$ is
known as B-field. In other words, by a slight abuse of notation the
parameters of the theory are
\be
\label{para} 
(\Lambda,B) \in \OO(2)\backslash \Gl(2)\times \Skew(2)\,.
\ee
Each $\Phi^\mu$ in (\ref{boac}) decomposes into a left- and a
rightmoving part $\Phi^\mu(z,\bar z) =$ $\frac{1}{2}(\varphi^\mu(z)+$
$\bar\varphi^\mu(\bar z))$, $\mu\in\{1,2\}$. The fields
$j_\mu=i\partial\varphi^\mu$ are the two generic abelian $\uu(1)$
currents of the theory which generate translations along the
coordinate axes of $\mathbb{T}^2$. The energy momentum tensor is given in the
Sugawara form
\be\label{emo}
T={1\over2}\left( :\!j_1 j_1\!: + :\!j_2 j_2\!:\right),\quad
\bar{T}={1\over2}\left( :\!\bar{\jmath}_1 \bar{\jmath}_1\!: 
+ :\!\bar{\jmath}_2 \bar{\jmath}_2\!:\right).
\ee
In the following, we will work with $\varphi^\mu$ and
$\bar\varphi^\mu$ separately, but the left-right transformed analogue
of some statement will often not be mentioned explicitly in order to
avoid tedious repetitions.

The Hilbert space $\mathcal{H}$ of our theory decomposes into an
infinite number of sectors according to different winding and momentum
numbers of the ground state. We label ground states with winding mode
$\lambda=m_2\lambda_1+m_1\lambda_2\in\Lambda$ and momentum mode
$\mu=n_2\mu_1+n_1\mu_2\in\Lambda^\ast$ by $| m_1,m_2,n_1,n_2\rangle$,
where $(\mu_1,\mu_2)$ is the basis dual to
$(\lambda_1,\lambda_2)$. With
\be
\label{charges} 
\left( p(\lambda,\mu), \bar{p}(\lambda,\mu) \right)
:= {1\over\sqrt2} \left( \mu-B\lambda+\lambda,
\mu-B\lambda-\lambda\right) 
\ee
and cocycle factors $c_{\lambda,\mu}$ the vertex operator
corresponding to $| m_1,m_2,n_1,n_2\rangle$ is
\be\label{vertex}
V_{\lambda,\mu} 
:= c_{\lambda,\mu}:\exp[ip(\lambda,\mu)\cdot\varphi(z)
+i\bar p(\lambda,\mu)\cdot\bar{\varphi}(\bar z)]:\,.
\ee
In particular, $V_{\lambda,\mu}$ has charge $\left( p(\lambda,\mu),
\bar{p}(\lambda,\mu) \right)$ with respect to
$(j_1,j_2,\bar{\jmath}_1,\bar{\jmath}_2)$, and by (\ref{emo}) the
action of the zero modes $\mbar{L}_0$ of the Virasoro generators is
$$
\mbar{L}_0|m_1,m_2, n_1, n_2\rangle 
=\frac{1}{2}\left(\mpbar{p}(\lambda,\mu)\right)^2
|m_1,m_2, n_1, n_2\rangle\,.
$$
Hence our theory has partition function 
\be
\label{Ztolambda}
Z_{\Lambda,B}(\sigma) = tr_{\mathcal{H}} {q^{L_0-c/24}\bar
{q}^{\bar{L}_0-c/24}} = \frac{1}{\eta^2\bar {\eta}^2} \sum_{\lambda\in
\Lambda,\mu\in \Lambda^\ast} q^{\frac{1}{2}(p(\lambda,\mu))^2}\bar
q^{\frac{1}{2}(\bar {p}(\lambda,\mu))^2}\,,
\ee
where $q = e^{2\pi i\sigma}$ and $\eta=\eta(\sigma)$ is the Dedekind
eta function
$$
\eta(\sigma) = q^{1/24}\prod_{n=1}^{\infty} (1- q^n) \,.
$$
By~\cite{cent,na} toroidal theories are determined uniquely by their
charge lattice
\be\label{latt}
\Gamma=\Gamma(\Lambda,B)
:=\left\{ (p(\lambda,\mu),\bar p(\lambda,\mu))\mid 
(\lambda,\mu)\in\Lambda\oplus\Lambda^\ast \right\}.
\ee
This is an even unimodular lattice in 
$\R^{2,2}=\R^2\times\R^2$ which is equipped with the scalar product
\be\label{skp}
(p,\bar p)\cdot (p^\prime, \bar p^\prime)
:= p\cdot  p^\prime - \bar p \cdot \bar p^\prime \,.
\ee
The parameters $(\Lambda,B)\in \OO(2)\backslash \Gl(2)\times \Skew(2)$
thus are mapped to $\Gamma(\Lambda,B)\in \OO(2)\times \OO(2)\backslash
\OO(2,2;\R)$.  The moduli space of toroidal conformal field theories
with central charge $c=2$ is
\be\label{narainmodulispace}
\mathcal{T}^2 = \OO(2)\times \OO(2)\big\backslash \OO(2,2;\R)
/\OO(2,2;\Z)
\ee
\cite{cent,na}. 
In the two-dimensional case it is convenient to group the four real 
parameters $G_{\mu\nu},B_{\mu\nu}$ of the theory
into two complex parameters by 
\be\label{taurho}
\tau = \tau_1 + i\tau_2 := \frac{G_{12}}{G_{22}} 
+ i\frac{\sqrt{\det(G_{\mu\nu})}}{G_{22}},
\quad
\rho = \rho_1 + i\rho_2 := B_{12} + i\sqrt{\det(G_{\mu\nu})}.
\ee
Here $\tau$ is the image of $\Lambda\in \Gl(2)$ under the natural
projection $\Gl(2)\rightarrow
\Sl(2)^{(\tau)}\cong\H=\{z\in\C\mid\Im(z)>0\}$.  If
$\OO(2,2;\R)\ni\Gamma(\Lambda,B)\mapsto (\tau,\rho)$, then
$\rho\in\H\cong \Sl(2)^{(\rho)}$, where $\Sl(2)^{(\rho)}$ is the
commutant of $\Sl(2)^{(\tau)}$ in $\OO(2,2;\R)$. Note that $\tau$ is
the quotient $\int_B dz/\int_A dz$ of the two torus periods ($A,B$
form a symplectic basis of $H_1(\mathbb{T}^2,\Z)$) and therefore represents the
complex structure of $\mathbb{T}^2$. $\rho_2$ is the volume of $\mathbb{T}^2$ and
specifies the K\a hler class, because $\dim_\R H^2(\mathbb{T}^2,\R)=1$ and
every metric on a two-dimensional torus is K\a hler. Therefore
$\rho\in\H$ is called complexified K\a hler parameter. Now the
generators $\lambda_1,\lambda_2\in\Lambda$ are given by
\be\label{bfield}
\lambda_1 = \sqrt{\rho_2\over\tau_2} \pmatrix {1\cr 0} ,\quad
\lambda_2 = \sqrt{\rho_2\over\tau_2} \pmatrix {\tau_1\cr\tau_2},
\qquad\mbox{and}\qquad
B={\rho_1\over\rho_2} \pmatrix{0&-1\cr 1&0} .
\ee
By (\ref{taurho}) for $\lambda=m_2\lambda_1+m_1\lambda_2\in\Lambda$
and $\mu=n_2\mu_1+n_1\mu_2\in\Lambda^\ast$ as above (\ref{charges})
reads
\be
\label{char}
\mpbar{p}
= {1\over\sqrt{2\tau_2\rho_2}} \left\{ 
\pmatrix{n_2\tau_2\cr -n_2\tau_1+n_1}
+ \rho_1 
\pmatrix{m_1\tau_2\cr -m_2-m_1\tau_1}
\pmm \rho_2 
\pmatrix{m_2+m_1\tau_1\cr m_1\tau_2} \right\}.
\ee
If $(\Lambda,B)$ are related to $(\tau,\rho)$ by (\ref{taurho}),
for the partition function (\ref{Ztolambda}) we write
\be
\label{Zto}
Z(\tau_1,\tau_2,\rho_1,\rho_2) := Z_{\Lambda,B}(\sigma) =
\frac{1}{\eta^2\bar {\eta}^2} \sum_{\lambda\in \Lambda,\mu\in
\Lambda^\ast} q^{\frac{1}{2}(p(\lambda,\mu))^2}\bar
q^{\frac{1}{2}(\bar {p}(\lambda,\mu))^2}\,.
\ee
Note that if $\tau_1=\rho_1=0$, then the torus theory is a tensor
product of two theories with $c=1$ corresponding to compactification
of single real bosons on circles of radii
$r=\sqrt{G_{22}}=\sqrt{\rho_2/\tau_2}$ and
$r^\prime=\sqrt{G_{11}}=\sqrt{\rho_2\tau_2}$.  The partition function
(\ref{Zto}) factorizes correspondingly:
\bea
Z^{c=1}(r)&:=& {1\over\left|\eta\right|^2}
\sum_{m,n\in\Z} q^{{1\over4}( {n/ r}+mr)^2}
\bar q^{{1\over4}( {n/ r}-mr )^2}\,,
\label{circlepart}\\
Z(0,\tau_2,0,\rho_2) &=& Z^{c=1}\left(\sqrt{\rho_2\over\tau_2}\right) 
Z^{c=1}(\sqrt{\rho_2\tau_2})\,.
\label{circle2}
\eea
In terms of the new parameters $(\tau,\rho)$ the duality group
$\OO(2,2;\Z)$ in (\ref{narainmodulispace}) translates into the group
generated by $\PSL(2,\Z)\times \PSL(2,\Z)$, which acts by M\o bius
transformations on each factor of $\H\times\H$, and the dualities
\be\label{UandV}
U,V: \H\times\H\rightarrow \H\times\H\,, \qquad
U(\tau,\rho) := (\rho,\tau), \qquad
V(\tau,\rho) := (-\bar\tau,-\bar\rho)\,.
\ee
In terms of the parameters $(\tau,\rho)$ the moduli space
(\ref{narainmodulispace}) therefore is
\be\label{modspace}
{\mathcal T}^2 = \left(\H/\PSL(2,\Z)\times\H/\PSL(2,\Z)\right) 
\big/ (\Z_2\times\Z_2).
\ee
By the above interpretation of $\tau$ and $\rho$ the duality $U$
interchanges complex and (complexified) K\a hler structure of $\mathbb{T}^2$
and is known as mirror symmetry. Compared to the former description
(\ref{narainmodulispace}) of the moduli space by equivalence classes
of lattices, $V$ correponds to conjugation by $\diag(-1,1,-1,1)$ on
$\OO(2)\times \OO(2)\backslash \OO(2,2;\R)$ which is target space
orientation change.  Note that world sheet parity which interchanges
$p$ and $\bar p$ is given by $(\Lambda,B)\mapsto(\Lambda,-B)$ or
equivalently $(\tau,\rho)\mapsto(\tau,-\bar\rho)$ and is not a duality
symmetry.

It is not hard to see that the Zamolodchikov metric on $\mathcal{T}^2$
is induced by the product of hyperbolic metrics on each of the factors
$\H$ in (\ref{modspace}).  In particular, geodesics on the Teichm\u
ller space $\H\times\H$ of $\mathcal{T}^2$ are well known: The
projection on each of the $\H$-factors is a half circle with center on
the real axis, a half line parallel to the imaginary axis of $\H$, or
constant.

Suppose that a nonisolated component of $\mathcal{C}^2$ with Teichm\u
ller space $\mathcal{E}\subset\H\times\H$ is obtained by modding out a
common symmetry group $G$ of all toroidal theories with parameters in
${\mathcal{E}}$. Assume further that $\mathcal{E}$ is a maximal
connected subset of $\H\times\H$ corresponding to theories with
symmetry $G$.  In particular, $G$ acts as group of isometries on
${\mathcal{E}}$, and the $(1,1)$-fields which describe deformations
within ${\mathcal{E}}$ are invariant under $G$. Thus ${\mathcal{E}}$
is totally geodesic.

Let us determine all possible actions of symmetry groups $G$ on
theories with parameters in ${\mathcal{E}}$.  Those best understood
are of course the ones with \emph{geometric interpretation}, i.e.\
those induced by an action of $G$ on the torus $\mathbb{T}^2=\R^2/\Lambda$ of a
geometric interpretation $(\Lambda,B)$.  We remark that if
$\mathcal{E}$ contains a \emph{large volume theory}, then the action
of $G$ does have a geometric interpretation.  Namely,
in~\cite[(1.16)]{nawe} a precise notion of large volume theories was
introduced, characterizing them by the fact that for the subset
\be\label{largevolume}
\widetilde{\Gamma}:=\Bigl\{(p,\bar p)\in\Gamma\mid 
\|p\|^2\ll1, \|\bar p\|^2 \ll1\Bigr\}
\ee
of the charge lattice $\Gamma$ the rank of
$\mbox{span}_\Z\widetilde{\Gamma}$ is two.  In particular, a large
volume theory has a unique preferred geometric interpretation
$(\Lambda, B)$ with large $\rho_2=\det(G_{\mu\nu})$ in terms of a
nonlinear $\sigma$ model.  If $\mathcal{E}$ contains such a large
volume theory with preferred geometric interpretation $(\Lambda,B)$,
then the action of $G$ on that toroidal theory will not change this
preferred geometric interpretation.  Since $\widetilde{\Gamma}$ as
defined in (\ref{largevolume}) has the property
$\mbox{span}_\Z\widetilde{\Gamma} =\{
{1\over\sqrt2}(\mu,\mu)\mid\mu\in\Lambda^\ast \}$, the action of $G$
is given by a geometric symmetry on the corresponding torus
$\mathbb{T}^2=\R^2/\Lambda$.

Let us assume that $G$ maps the set $\{j_k\bar\jmath_l\mid
k,l\in\{1,2\}\}$ of generic $(1,1)$ fields of theories in
$\mathcal{T}^2$ into itself.  By construction of toroidal conformal
field theories, this means that $G$ induces an action on the entire
Teichm\u ller space $\H\times\H$ of $\mathcal{T}^2$, which identifies
isomorphic theories and fixes $\mathcal{E}$. This action will be
denoted $\mathcal{G}$ in the following.  By construction
(\ref{modspace}) of the moduli space $\mathcal{T}^2$ of toroidal
theories, we must have $\mathcal{G}\subset \PSL(2,\Z)^2\rtimes\Z_2^2$.
Note that in general $\mathcal{G}$ will be different from $G$, since
an action of $G$ on vertex operators~(\ref{vertex}) by multiplication
with phases will be invisible in its induced action on
$\mathcal{T}^2$.  Moreover, target space orientation change $V$
induces a trivial action on toroidal conformal field theories.

If $\mathcal{E}$ contains a geodesic with the property that its
projection on one of the factors of $\H$ in the Teichm\u ller space is
constant, then by~(\ref{char}) one checks that $\mathcal{E}$ contains
a large volume limit and thus $G$ acts geometrically by the
above. Otherwise, since $\mathcal{E}$ is the fixed point set of a
subgroup $\mathcal{G}\subset \PSL(2,\Z)^2\times\Z_2^2$, and the action
of $V$ need not be discussed, $\mathcal{E}$ must be one of the
following spaces (or a M\o bius transform thereof):
\be
\label{excomp}
\mathcal{E}_U :=\left\{ (\tau_1,t,\tau_1,t)\mid t\in\R^+\right\},
\qquad \mathcal{E}_{UV}:=\left\{ (\tau_1,t,-\tau_1,t)\mid
t\in\R^+\right\}.
\ee
We now argue that in neither of these cases we find new components of
$\mathcal{C}^2$ by modding out a nongeometric symmetry.  Firstly,
since $\mathcal{E}$ is maximal, we may assume $G=\{1,g\}$, where
$g\in\{U,UV\}$ and $\mathcal{E}=\mathcal{E}_g$.  Then, for mirror
symmetry $g=U$ we read off an induced action $n_2\leftrightarrow m_2$
on the charge lattice~(\ref{char}).  Moreover, all theories in
$\mathcal{E}_U$ have a righthanded $\SU(2)\times \U(1)$ symmetry, two of
whose commuting generators are invariant under this action. Since one
\pagebreak[3]
checks that all the generic abelian lefthanded $\U(1)$ currents are
invariant under the action of $U$ as well, we find that the theory we
produce by modding out $U$ contains at least two left- and two
righthanded abelian currents and thus is a torus theory again. $U$
therefore is $\SU(2)$ conjugate to a shift on the charge lattice, which
acts by multiplication with $i^{n_2-m_2}$ on states created from the
Hilbert space ground state $|m_1,m_2,n_1,n_2\rangle$. It is now a
straightforward calculation to check that performing this shift
orbifold reproduces the original theory.  The case $g=UV$ is treated
analogously, since $\mathcal{E}_{UV}$ is obtained from $\mathcal{E}_U$
by a parity change $(\tau,\rho)\mapsto(\tau,-\bar\rho)$.

Summarizing, up to now we have shown that if the set
$\{j_k\bar\jmath_l\mid k,l\in\{1,2\}\}$ of generic $(1,1)$ fields of
theories with parameters in $\mathcal{E}$ is mapped onto itself by the
action of $G$, then this action possesses a geometric
interpretation. Otherwise we call the action as well as the
corresponding orbifold component of $\mathcal{C}^2$
\emph{exceptional}. In fact, since the Teichm\u ller space
$\mathcal{E}$ of an exceptional component is totally geodesic, to give
an estimate of how many exceptional components one may find it
suffices to determine all geodesics in $\H\times\H$ that parametrize
theories which generically possess more than four $(1,1)$ fields.  By
explicit calculation using~(\ref{char}) one checks that all such
geodesics have the form $f(t)=(\tau_1,t,\pm\tau_1,t)\in\H\times\H,
t\in\R^+$, or are M\o bius transforms thereof. In other words, without
loss of generality $\mathcal{E}=\mathcal{E}_U$ or
$\mathcal{E}=\mathcal{E}_{UV}$ as defined in~(\ref{excomp}).  Thus in
all exceptional cases the toroidal conformal field theories with
parameters in $\mathcal{E}$ possess an additional left- or righthanded
$\SU(2)$ symmetry, and the exceptional action is given by a binary
tetrahedral, octahedral or icosahedral subgroup $T,O,I$ of $\SU(2)$
(see~\cite{par}), possibly in combination with some other symmetry.
For instance, if $\tau_1=0$ the toroidal theories in
$\mathcal{E}_U=\mathcal{E}_{UV}$ decompose into tensor products of
$c=1$ circle theories at radii $r=1$, $r^\prime=t$,
respectively~(\ref{circle2}). Then the possible actions of $T,O,I$ on
the first factor theory are clear from the results on conformal field
theories with central charge $c=1$~\cite{par}.  In general,
exceptional components of $\mathcal{C}^2$ are an interesting issue to
be studied separately, which exceeds the scope of the present paper.

We rather concentrate on the nonexceptional components of
$\mathcal{C}^2$ in the following.  Note that equivalent toroidal
theories need not always be mapped onto equivalent orbifold theories
if we mod out a symmetry group $G$, since the action of $G$ in some
cases does depend on the particular choice of coordinates on $\mathbb{T}^2$.
In other words, $\mathcal{C}$ is obtained from $\mathcal{E}$ by
modding out a subgroup of $\smash{\{A\in \PSL(2,\Z)^2\rtimes\Z^2\mid
A{\mathcal{E}}={\mathcal{E}}\}}$ which needs to be determined for every
group $G$ separately.

Recall on the other hand that every theory that was constructed as
orbifold by a solvable group $G$ possesses a symmetry which one can
mod out to regain the original theory \cite[section~8.5]{gin}. In
section~\ref{symm} we will see that indeed only orbifolds by solvable
groups are of relevance to us.  Thus no information distinguishing two
theories may be lost under our orbifold procedures. In other words, if
we mod out two distinct toroidal theories by the same symmetry, then
the resulting theories must be distinct as well.

\section{Symmetries of the two-dimensional torus}\label{symm}

By the discussion in section~\ref{tori}, to find the nonexceptional
nonisolated orbifold components of the moduli space $\mathcal{C}^2$ we
must employ the orbifold procedure for all possible discrete symmetry
groups of the torus.  In two dimensions, there are seventeen
inequivalent crystallographic space groups~\cite{orgcrystal}, i.e.\
discrete subgroups $G \subset \OO(2) \ltimes \R^2$ that leave
invariant some lattice $\Lambda^\prime$ and therefore act on a torus
$\mathbb{T}^2 = \R^2/\Lambda$, where $\Lambda\subset\Lambda^\prime$.
Figure~\ref{lattice}
shows all these symmetry groups by depicting the orbit of some symbol
$\blacktriangleright$ under $G$.  Each lattice $\Lambda^\prime$ in
figure~\ref{lattice} is formed by fixed combinations of the symbol
$\blacktriangleright$, which we call motive, in various orientations.
Then $\Lambda\subset\Lambda^\prime$ is given by those motives which
have the same orientation.  The space group G is a semi-direct product
of a finite point group $P\subset \OO(2)$ and a ``translationary''
group $\triangle\subset \OO(2)\ltimes\R^2$ of elements which do not
fix the origin.  In figure~\ref{lattice} the group $\Delta$ is the
minimal subgroup of $G$ which acts transitively on motives.  The
finite group $P$ is determined by inspection of the particular motives
which comprise the orbit of the symbol $\blacktriangleright$ under $P$
each.

By the above, P is an automorphism group of the two-dimensional
lattice $\Lambda$, and if $(S,\delta)\in\Delta$, then there is some $N
\in \Z$ such that $N \delta\in\Lambda$.  Therefore if $A\in P$ has
order M then $M \in \{2,3,4,6\}$. The values $M=3$ or $M=6$ require
$\Lambda$ to be a hexagonal lattice ($\tau=e^{2\pi i/3}$); $M = 4$
requires a square lattice ($\tau=i$).  As to symmetry groups of order
$M=2$, $\Z_2$ acts by $x \mapsto -x$ as automorphism on every lattice
$\Lambda$. Moreover, the reflection symmetry group $\Z_2(R)$ is an
automorphism group of lattices with $\tau_1\in\{0,{1}/{2}\}$,
where R acts on the coordinates of $\mathbb{T}^2$ by
\be\label{refl}
R = R_1: (x^1,x^2) \mapsto (x^1, -x^2)\qquad\mbox{or}\qquad
R = R_2: (x^1,x^2) \mapsto (-x^1, x^2) \,.
\ee
Translations $T_{\delta} = e^{2\pi i p\cdot{\delta\over\sqrt2}}$ by
$\delta \in \Lambda$ are the basic symmetries of the torus $\mathbb{T}^2 =
\R^2/\Lambda$.  The result of modding out any torus by a translation
symmetry $T_{\delta}$, $N \delta \in \Lambda$, $N\in \N$ minimal with
this property, gives another torus with lattice generated by $\Lambda$
and $\delta$.  To produce a surface different from the torus (and
later on non-toroidal conformal field theories), we must combine the
translation with the reflection symmetry which we denote $T_R := R
e^{2\pi i p\cdot{\delta\over\sqrt2}}$. More precisely, we will need
this symmetry only in the case $\tau_1=0$ and $N=2$, and we set
\be
\label{symmetries}
\begin{array}{rclrclrcl}
\delta_1 &:=&\ds \sqrt{\rho_2\over \tau_2}\!\pmatrix{1/2\cr 0}, &
\delta_2 &:=&\ds \sqrt{\rho_2\over \tau_2}\!\pmatrix{0\cr\tau_2/2}, &
\delta^\prime &:=&\ds \sqrt{\rho_2\over \tau_2}\!\pmatrix{1/2\cr\tau_2/2}\!;
\\[10pt]
\mbox{for }\mu\in\{1,2\}:\; T_{R_\mu} &:=& R_\mu e^{2\pi i
p\cdot{\delta_\mu\over\sqrt2}}, &T_{R_\mu}^\prime &:=& R_\mu e^{2\pi i
p\cdot{\delta^\prime\over\sqrt2}}, &\widehat{T}_{R_2} &:=& R_2 e^{2\pi
i p\cdot{\delta_1\over\sqrt2}}.
\end{array}
\ee 
The groups of type $\Z_2$ generated by $T_R$ or $T_R^\prime$ are
denoted $\Z_2(T_R)$ or $\Z_2(\mathbb{T}^\prime_{R})$, respectively, where
either $R=R_1$ or $R=R_2$.  We denote by $A(\theta) \in \Z_M$ the
rotation\linebreak\pagebreak[3]
\FIGURE{
\hspace*{\fill}
\leavevmode
\epsfxsize=0.8\textwidth
\epsffile{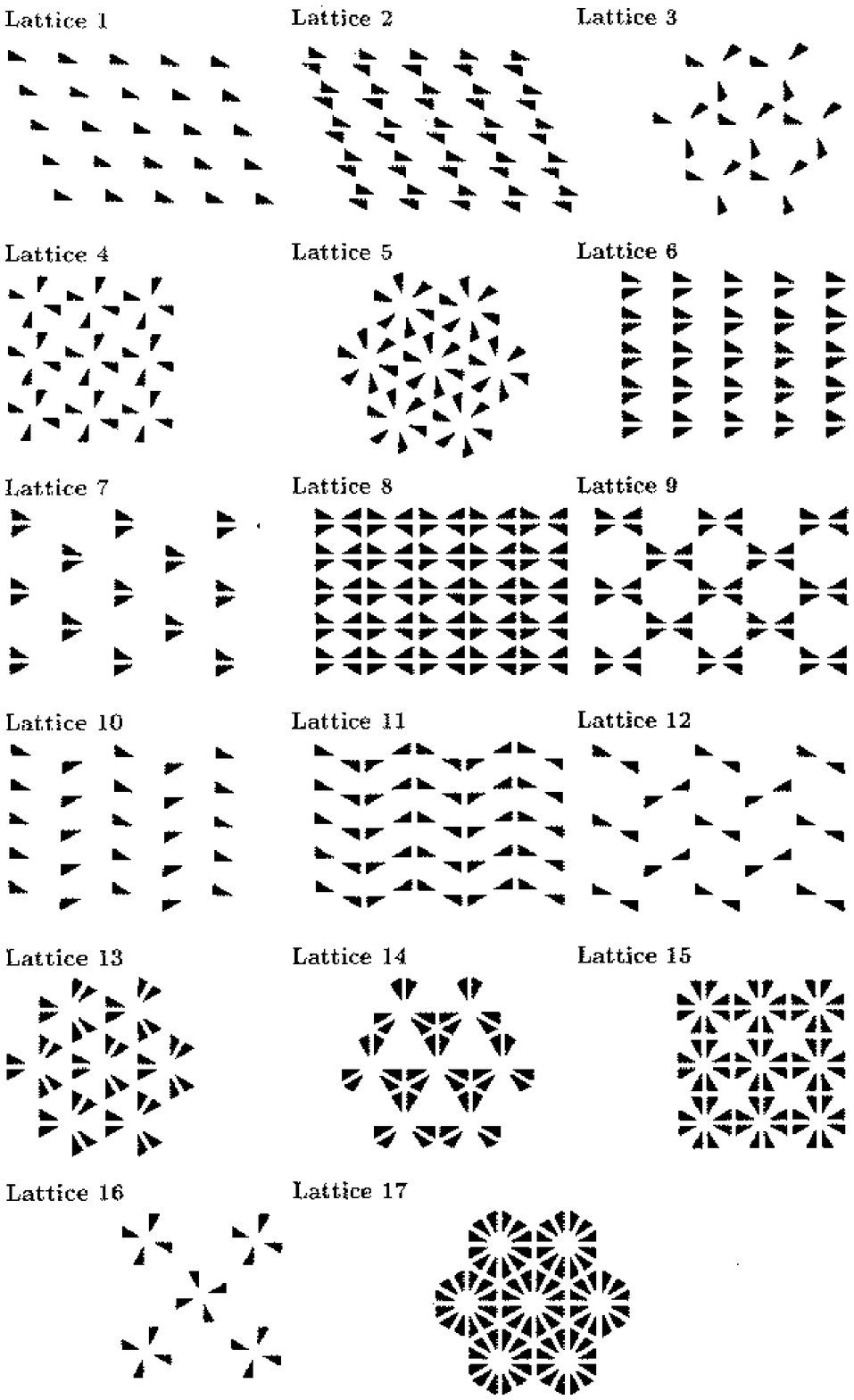}\hspace*{\fill}
\caption{The seventeen inequivalent crystallographic space 
groups \cite{crystal}.}\label{lattice}
}
\clearpage
\noindent by an angle of $\theta$. Then $R_2 = A(\pi)R_1$,
$\mathbb{T}^{(\prime)}_{R_2} = A(\pi)\mathbb{T}^{(\prime)}_{R_1}$, and we have the
noncyclic crystallographic groups
$$
\begin{array}{rclrcl}
D_2 & := & \{ 1, A(\pi), R_1, R_2 \}\,,
&D_3(R) &:= & \Z_3 \cup R\Z_3\,,
\\[3pt]
D_4 &:= &\Z_4 \cup R_1\Z_4 = \Z_4 \cup R_2\Z_4\,,\qquad
&D_6 &:=& \Z_6\cup R_1\Z_6 = \Z_6\cup R_2\Z_6\,,
\\[3pt]
D_2(T_R) &:= & \{1, A(\pi), T_{R_1}, \widehat{T}_{R_2} \}\,,
&D_2(\mathbb{T}^{\prime}_R) 
&:=&  \{1, A(\pi), \mathbb{T}^{\prime}_{R_1}, \mathbb{T}^{\prime}_{R_2} \}\,,
\\[3pt]
D_4(\mathbb{T}^{\prime}_R) &:=& \Z_4 \cup \mathbb{T}^{\prime}_{R_1} \Z_4 = 
\Z_4 \cup \mathbb{T}^{\prime}_{R_2} \Z_4\,.
\end{array}
$$
The symmetries that correspond to the lattices in figure~\ref{lattice} are
\be\label{group}
\begin{array}{l}
\begin{array}{l||c|c|c|c|c|c|c|c|c}
\mbox{Lattice} 
& 1 & 2 & 3 & 4 & 5 & 6 & 7 & 8 & 9\\ \hline
\mbox{Symmetries} 
& \{T_{\delta},\delta \in \Lambda \} & \Z_2 & \Z_3 & \Z_4 
              & \Z_6 & \Z_2(R) & \Z_2(R) & D_2  & D_2\\
\end{array} \\[20pt]
\begin{array}{l||c|c|c|c|c|c|c|c}
\mbox{Lattice}     
& 10 & 11 & 12 & 13 & 14 & 15 & 16 & 17 \\\hline
\mbox{Symmetries}   
& \Z_2(T_{R}) & D_2(T_{R}) & D_2(\mathbb{T}^\prime_{R}) & 
       D_3(R_1) & D_3(R_2) & D_4 & D_4(\mathbb{T}^\prime_R) & D_6 \\
\end{array}.
\hspace*{5pt}
\end{array}
\ee
Note that as anticipated at the end of section~\ref{tori}, all groups
occuring in~(\ref{group}) are solvable. This is clear for the abelian
groups.  For the dihedral groups $D_n$ it follows from the fact that
the subgroup $\Z_n$ of rotations in $D_n$ is a normal subgroup with
abelian factor $D_n/\Z_n\cong\Z_2$.  The finite reflection groups
among the groups listed in~(\ref{group}) are $\Z_2(R)$, $D_2$,
$D_3(R)$, $D_4$, and $D_6$. These are better known as Weyl groups of
the semisimple Lie algebras $A_1, A_1\oplus A_1, A_2, B_2$, and $G_2$,
respectively.
\vspace{-.3em}

\section{Sixteen orbifolds of the torus theories with $c=2$}\label{part}
\vspace{-.1em}

\looseness=-1 In~(\ref{group}) we have listed all the seventeen possible symmetry
groups $G$ of a two-dimensional torus $\mathbb{T}^2=\R^2/\Lambda$.  Because the
first of them, corresponding to lattice 1, is the translation group $G
\cong \Lambda$ which acts trivially on $\mathbb{T}^2$, this implies that we can
construct at most sixteen different types of orbifold theories
corresponding to different compactifications on $\mathbb{T}^2/G$.  To do so, we
must show that these symmetries can be continued to symmetries of the
corresponding two-dimensional conformal field theories.  Since the
action of $g\in G$ on the abelian currents $j_\mu$, which generate
translations along the coordinate axes of $\mathbb{T}^2$, is determined by the
action on $\mathbb{T}^2$, this amounts to continuing every $g\in G$ to a
symmetry of the charge lattice~(\ref{latt}).  By~(\ref{char}) it is
easy to see that this is possible iff $B=g^TBg$.  In particular, any
of the symmetries listed in~(\ref{group}) which corresponds to a
lattice characterized by parameters $\tau\in\H$ and $\rho_2\in\R^+$
immediately gives a symmetry of the toroidal conformal field theory
with parameters $(\tau,0,\rho_2)$, i.e.\ $B=0$.  But nonzero values
for $\rho_1$ might be possible, too.  Note in particular that
$\rho_1$, as parameter in $\mathcal{T}^2$, is only defined modulo
$\Z$. In other words, $g$ can be continued to a symmetry of the
toroidal conformal field theory iff 
\be
\label{cont} 
B= g^TBg + {n\over\rho_2} \pmatrix{0&-1\cr 1&0}, \qquad n\in\Z\,.
\ee 
Below, we will discuss all possible B-field values for each of the
symmetry groups listed in~(\ref{group}).
\pagebreak[3]

Let us recall how we can construct new conformal field theories by
modding out a symmetry group G of a conformal field theory with
central charge $c$ (see also~\cite{dixon, Dix,har,vafa,frva87}).
First we must project onto group invariant states in the Hilbert space
$\mathcal{H}$ of our theory to obtain the untwisted sector of the new
theory.  In the operator formalism this is achieved by the projection
operator \smash{$P:= \frac{1}{|G|}\sum_{g\in G}\limits g$}. We
employ the shorthand notation
$$
{\scriptstyle{g}}\square_{1}\limits \;:= 
\tr_{\mathcal H} g q^{L_0 -\frac{c}{24}}\bar{q}^{\bar L_0 -\frac{c}{24}}
$$
to write the 
untwisted sector partition function  as 
\be\label{gun}
Z_u =\tr_{\mathcal H} P q^{L_0 -\frac{c}{24}}\bar{q}^{\bar L_0 -\frac{c}{24}}
     = \frac{1}{|G|}\sum_{g\in G}{\scriptstyle{g}}\square_{1}.
\ee
$Z_u$ is not modular invariant. The reason is that the Hilbert space
of the new theory will also contain twisted sectors ${\mathcal H}_f,
f\in G$, corresponding to fields which are only well defined on the
world sheet of the original theory up to the action of a nontrivial
element $f\in G$:
\be\label{trans}
|\varphi\rangle \in {\mathcal H}_f\,: \qquad
\varphi(\xi_0,\xi_1+1) = f \varphi(\xi_0,\xi_1)\, .
\ee
More precisely, we should label twisted sectors by conjugacy classes
$\{f\}$ of $G$ because $\varphi$ as in~(\ref{trans}) also obeys
\be\label{conj}
g\varphi(\xi_0,\xi_1+1) = (gfg^{-1})g \varphi(\xi_0,\xi_1)
\ee
for any $g\in G$, and $g\varphi$ is identified with $\varphi$, so $\fa
g\in G:\,\mathcal{H}_f=\mathcal{H}_{gfg^{-1}}$.  $\mathcal{H}^G$ acts by the
induced representation on the entire twisted sector.

For $|\varphi\rangle\in {\mathcal H}_f, f\neq1$ by~(\ref{trans}) we
find that $q_j:= \varphi(z=0)$ is a fixed point of $f$. If $f$ has $J$
fixed points on $\mathbb{T}^2$ then ${\mathcal H}_f$ decomposes into $J$
isomorphic copies of spaces ${\mathcal H}_{f}^{(j)}$, $j\in\{ 1,
\dots, J\}$. If $f$ has order M, then $\varphi^\pm:=\varphi^1\pm
i\varphi^2$ has mode expansion
\be\label{zmmod}
\varphi^\pm (z) = q^\pm_j + i\sum_{n\in \Z \pm 1/M}\frac{1}{n} 
\alpha_n^\pm z^{n} \,,
\ee
so the corresponding 
twisted ground state has dimensions 
\be\label{dim}
h = \bar h = \frac{1}{2} \frac{1}{M}\left(1-\frac{1}{M}\right) 
\ee
(see also~\cite{Dix}).  In the twisted Hilbert space $\mathcal{H}_f$,
we again have to project onto group invariant states, now by $P_f:=
\frac{1}{|G|}\sum_{g\in G:\, [g,f]=0}\limits g$.  The prefactor is
adjusted correctly in order to take care of the multiplicities in each
twisted sector. Namely, it takes care of overcounting if later on we
sum over all $f\in G, f\neq1$ instead of conjugacy classes $\{f\}$
which actually label twisted sectors by the above (see~\cite{Dix}).
We again use the shorthand
$$
{\scriptstyle{g}}\square_{f}\limits \;:= \tr_{{\mathcal H}_f} gq^{L_0
-\frac{c}{24}}\bar{q}^{\bar L_0 -\frac{c}{24}}
$$
to write the twisted sector partition function  as 
\be
\label{gtiw}
Z_t =\sum_{f\in G, f\neq1}\limits \tr_{{\mathcal H}_f} P_f q^{L_0
-\frac{c}{24}}\bar{q}^{\bar L_0 -\frac{c}{24}} =
\frac{1}{|G|}\sum_{g,f\in G,\atop f\ne 1,
[g,f]=0}{\scriptstyle{g}}\square_{f}\,.
\ee
The total modular invariant orbifold partition function is 
\be
\label{gorb}
Z_{G-{\rm orb}} = \sum_{f\in G} \tr_{{\mathcal H}_f} P_f q^{L_0
-\frac{c}{24}} \bar{q}^{\bar L_0 -\frac{c}{24}} =
\frac{1}{|G|}\sum_{g,f\in G, \atop
[g,f]=0}{\scriptstyle{g}}\square_{f} \,,
\ee
where we set ${\mathcal H}_1\;:= \mathcal H$ and $P_1\;:= P$.
For general $f,g\in G$ the contribution  
$\smash{{\scriptstyle{g}}\square_{f}\limits}$ 
can also be calculated by using modular transformations:
\be
\label{trafo}
{\scriptstyle{g}}\square_{f}\left(-\frac{1}{\sigma}\right) =
{\scriptstyle{f}}\square_{g}(\sigma)\,,\qquad
{\scriptstyle{g}}\square_{f}(\sigma + 1) =
{\scriptstyle{f\circ g}}\square_{f}(\sigma)\,.
\ee
Note that ${\scriptstyle{g}}\square_{f}\limits$ a priori is only
defined up to a phase, because the same is true for the action of
$g\in G$ on a twisted ground state of $\mathcal{H}_f$.  Only if
$g=f^k$ for some $k\in\Z$, the phase is fixed by~(\ref{trafo}), and
for all other boxes the choice is restricted by modular
invariance. For closed modular orbits in the twisted sector there
remains an arbitrariness of the phase they contribute with. Here,
conjugate subgroups must account with the same phase in order for the
representation of $G$ on the twisted sector to be consistent
with~(\ref{conj}).  This ambiguity, which by the above does not occur
for orbifolds by cyclic groups, is known as discrete torsion
\cite{cum} and will become relevant in the discussion of lattices 8
and 9 as well as 15--17 below.  Because the only groups this will occur
in are of type $D_2$, discrete torsion in these cases will always be
given by a choice of sign only.

For nonabelian G, (\ref{gorb}) can be written as sum over abelian
subgroups of G with overcounted terms subtracted off. To do so, we
call a subgroup $H\subsetneqq G$ maximal abelian if there is no
abelian $G'\subsetneqq G$ such that $H\subsetneqq G'$. We also
introduce multiplicities $n_{H'}:= \#\{ H\subsetneqq G \mbox{
maximal abelian }| H'\subsetneqq H \mbox{ maximal}\}$ and find
\be
\label{nabe}
Z_{G-{\rm orb}} = \frac{1}{|G|}\left( \sum_{ H\subsetneqq G \mbox{
\scriptsize max.} \atop \mbox{\scriptsize abelian }} |H| Z_{H-{\rm orb}} -
\!\sum_{H'\subseteqq G:\,\exists H\subsetneqq G\atop \mbox{ \scriptsize
max. abelian, } H'\subsetneqq H} (n_{H'}-1) Z_{H'-{\rm orb}} \right).
\ee

\subsection{Lattices 2 to 5: $\Z_M$ orbifold theories}

We briefly describe the $\Z_M$ orbifold construction.  For details
see~\cite{zm}, where the $\Z_M$ orbifold partition functions were
constructed for $c=3$ superconformal field theories. Most of the
arguments translate directly to the purely bosonic case with $c=2$
studied here.

In the following, let $\gamma$ be a generator of $\Z_M$ and assume
$\mathbb{T}^2 =\R^2/\Lambda$ to be a torus with $\Z_M$ symmetry, where
$\Lambda$ is characterized by specific values of $\tau$ and $\rho_2$
as explained in section~\ref{tori}. By the discussion in
section~\ref{symm} this means that $\tau=e^{2\pi i/M}$ if
$M\in\{3,4,6\}$, and $\rho_2\in\R^+$ arbitrary, whereas $\Z_2$ is a
symmetry for every torus.  Because by~(\ref{bfield}) $\gamma$ commutes
with $B$ for any value of $\rho_1$, from~(\ref{cont}) we know that
every toroidal conformal field theory with parameters $(\tau,\rho)\in
\H\times\H\,$, $\tau=e^{2\pi i/M}$ for $M\in\{3,4,6\}$, has $\Z_M$
symmetry.  The action of the rotation group $\Z_M$ on the charge
lattice~(\ref{latt}) is given by
\be\label{act}
\gamma\in \Z_M\,,\qquad \gamma : (p,\bar p) 
\mapsto   (\gamma p, \gamma \bar p)\,.
\ee
It follows that the $\Z_M$ action commutes with M\o bius
transformations on $\rho$.  The $\Z_2$ action commutes with the entire
$\PSL(2,\Z)^2\rtimes\Z^2_2$ of~(\ref{modspace}), so for the families of
$\Z_M$ orbifold conformal field theories with $c=2$ we get the
following irreducible components of $\mathcal{C}^2$:
\bea\label{zmpara}
\mathcal{C}_{\Z_2-{\rm orb}}&\cong& \mathcal{T}^2\,,
\nonumber\\
\mbox{for }M\in\{3,4,6\}:\quad \mathcal{C}_{\Z_M-{\rm orb}} &=&
\left\{ (\tau,\rho)\mid \tau=e^{2\pi i/M},
\rho\in\H/\PSL(2,\Z)\right\}
\nonumber\\
&\cong& \;\H/\PSL(2,\Z)\,.
\eea
By~(\ref{act}) the Hilbert space sectors built on the ground states
$|m_1,m_2,n_1,n_2\rangle$ are permuted by the $\Z_M$ action, the only
fixed ground state being $|0,0,0,0\rangle$. Since the
$|m_1,m_2,n_1,n_2\rangle$ are pairwise orthogonal, the only
contribution to
$\sum_{k=1}^{M-1}{\scriptstyle{\gamma^k}}\square_{1}\limits$
in~(\ref{gun}) comes from the Hilbert space sector built on
$|0,0,0,0\rangle$.  The $\Z_M$ action on oscillator modes is read off
from
\be
(\gamma^k\varphi^{\pm})(z) = e^{\pm \frac{2\pi i k}{M}} \varphi^{\pm}(z)\,,
\qquad k \in\{ 1,2,\dots, M-1\} \,.
\ee
This allows to construct the untwisted sector partition function
(\ref{gun}). The twisted sector partition function~(\ref{gtiw}) is
either obtained by using~(\ref{dim}) and~(\ref{zmmod}) to calculate
every box ${\scriptstyle{\gamma^k}}\square_{\gamma^l}\limits\,,
l\neq0,$ separately or by modular transformations.

\paragraph{Lattice 2: the $\Z_2$ orbifold.}

By~(\ref{zmpara}) lattice 2 depicts an arbitrary lattice.  (\ref{gun})
and the above show that for any $(\tau,\rho)\in\H\times\H$
$$
Z_u = \frac{1}{2}\left(Z(\tau,\rho) + \frac { (q\bar
q)^{-{1}/{12}}} {\prod_{n=1}^{\infty}(1 + q^n)^2(1 +\bar{q}^n)^2}
\right) = \frac{1}{2} \left( Z(\tau,\rho) + 4\left|
\frac{\eta(\sigma)} { \vartheta_2(\sigma) } \right|^2 \right) .
$$
Here and in the following $\vartheta_i(y,\sigma)$, $i \in\{1,
\dots,4\}$ denote the classical Jacobi theta functions, and
$\vartheta_i(\sigma):= \vartheta_i(0,\sigma)$.
\pagebreak[3]

Every torus $\mathbb{T}^2$ has four fixed points under the $\Z_2$
symmetry.  By~(\ref{dim}) this yields four twisted ground states with
conformal dimensions $(h, \bar h) =
({1}/{8},{1}/{8})$. By~(\ref{gtiw}) we find for the twisted
sector partition function
\bea
\label{z2tiw}
Z_{t} &=& 4\cdot\frac{1}{2} (q\bar q)^{-{1}/{12}}\left(\left|
q^{{1}/{8}} \prod_{n=1}^{\infty}(1 - q^{n-1/2})^{-2}\right|^2 
+ \left| q^{{1}/{8}} \prod_{n=1}^{\infty}(1 +
q^{n-1/2})^{-2}\right|^2 \right)
\nonumber\\
&=& 4\cdot \frac{1}{2} \left( \left| \frac{\eta(\sigma)} {
\vartheta_4(\sigma) }\right|^2 + \left| \frac{\eta(\sigma)}{
\vartheta_3(\sigma) }\right|^2 \right) .
\eea
The complete $\Z_2$ orbifold partition function is 
\be
\label{lattice2}
Z_{\Z_2-{\rm orb}} (\tau,\rho) = \frac{1}{2} \left(Z(\tau,\rho) +
4\left| \frac{\eta(\sigma)}{\vartheta_2(\sigma)} \right|^2  + 
4\left| \frac {\eta(\sigma)}{\vartheta_4(\sigma) } \right|^2  + 
4\left| \frac {\eta(\sigma)}{\vartheta_3(\sigma) } \right|^2 \right) .
\ee
The analogous formula for  $\Z_2$ orbifold conformal field theories
with $c=1$ is of course also well known~\cite{egrs87,sa87,ya87}:
\be
\label{c1orb}
Z_{\rm orb}^{c=1} (r) = \frac{1}{2} \left(Z^{c=1} (r) + 2\left|
\frac{\eta(\sigma)}{\vartheta_2(\sigma)} \right|  +  2\left| \frac
{\eta(\sigma)}{\vartheta_4(\sigma) } \right| +  2\left| \frac
{\eta(\sigma)}{\vartheta_3(\sigma) } \right| \right) ,
\ee
where $Z^{c=1}(r)$ was given in~(\ref{circlepart}).

\paragraph{Lattice 3: the $\Z_3$ orbifold.}

Lattice 3 has $\tau = e^{2\pi i/3}$ and by~(\ref{zmpara}) we may
pick  arbitrary $\rho\in\H/\PSL(2,\Z)$. 
The untwisted sector partition function  is 
$$
Z_u = \frac{1}{3}\left(Z(\tau = e^{2\pi i/3},\rho) + 6
\left|\frac{\eta(\sigma)} {\vartheta_1(\frac{1}{3},\sigma)}\right|^2
\right) .
$$
$\Z_3$ symmetric tori have three fixed points under the $\Z_3$ action.
Thus by~(\ref{dim}) there are three twisted ground states of
dimensions $(h,\bar h) = (1/9,1/9)$ in each of the twisted sectors
${\mathcal H}_{\gamma}$ and ${\mathcal H}_{\gamma^2}$.  The twisted
sector partition function therefore is
\be
\label{z3tiw}
Z_{t} =2\cdot3\cdot\frac{1}{3}(q\bar{q})^{-{1}/{18}}\sum_{l=1}^{3}
\left|\frac{\eta(\sigma)} {\vartheta_1( \frac{\sigma}{3}
+\frac{l}{3},\sigma)}\right|^2 ,
\ee
and for the complete $\Z_3$ orbifold partition function,
\be
\label{lattice3}
Z_{\Z_3-{\rm orb}}(\tau = e^{2\pi i/3}, \rho) = \frac{1}{3}\left( Z +
6 \left|\frac{\eta(\sigma)} {\vartheta_1(\frac{1}{3},\sigma)}\right|^2
+ 6(q\bar{q})^{-{1}/{18}}\sum_{l=1}^{3} \left|\frac{\eta(\sigma)}
{\vartheta_1( \frac{\sigma}{3} +\frac{l}{3},\sigma)}\right|^2\right) .
\ee

\paragraph{Lattice 4: the $\Z_4$ orbifold.}

Lattice 4 has $\tau = i$, and by~(\ref{zmpara}) we may pick arbitrary
$\rho\in \H/\PSL(2,\Z)$.  The untwisted sector partition function of
the $\Z_4$ orbifold can be written as
$$
Z_u = \frac{1}{4} \left( Z(\tau = i, \rho) +
4\left|\frac{\eta(\sigma)}{\vartheta_1(\frac{1}{4},\sigma)}\right|^2 +
4\left|\frac{\eta(\sigma)}{\vartheta_2(\sigma)}\right|^2 \right) .
$$
Tori with $\tau=i$ have three fixed points under the rotation group
$\Z_4$, one of which corresponds to a $\Z_2$ twist and two to $\Z_4$
twists.  Hence the total $\Z_4$ orbifold partition function is the sum
of untwisted, $\Z_2$, and $\Z_4$ twisted sector partition functions
$$
Z_{\Z_4-{\rm orb}}(\tau = i, \rho) = Z_u + Z_{2t} + Z_{4t} \,.
$$
The $\Z_2$ twisted sector partition function $Z_{2t}$ can be read off
from~(\ref{z2tiw}) by omitting the factor of four.  By~(\ref{dim}),
the two ground states in each of the twisted sectors ${\mathcal
H}_{\gamma}$, ${\mathcal H}_{\gamma^2}$ have dimensions $(h, \bar h) =
(3/32,3/32)$. The $\Z_4$ twisted sector partition function therefore
is
$$
Z_{4t} = \frac{1}{4}\left(
4\left|\frac{\eta(\sigma)}{\vartheta_4(\frac{1}{4},\sigma)}\right|^2 +
2\left|\frac{\eta(\sigma)}{\vartheta_3(\sigma)}\right|^2 +
2\left|\frac{\eta(\sigma)}{\vartheta_4(\sigma)}\right|^2 +
4(q\bar{q})^{-\frac{1}{32}}\sum_{l=1}^{4} \left|\frac{\eta(\sigma)}
{\vartheta_1( \frac{\sigma}{4} +\frac{l}{4},\sigma)}\right|^2\right) .
$$
Altogether, we find
\bea
\label{lattice4}
Z_{\Z_4-{\rm orb}}(\tau = i, \rho) & = &\frac{1}{4}\left(Z(\tau =i,
\rho) + 4\sum_{i=2}^{4}\left|\frac{\eta(\sigma)}
{\vartheta_i(\sigma)}\right|^2 + 4\left|\frac{\eta(\sigma)}
{\vartheta_1(\frac{1}{4},\sigma)}\right|^2+\right. 
\nonumber\\ & & 
	\hphantom{\frac{1}{4}\Biggl(}\!
+ \left.  4 \left|\frac{\eta(\sigma)}
{\vartheta_4(\frac{1}{4},\sigma)}\right|^2 +
4(q\bar{q})^{-{1}/{32}}\sum_{l=1}^{4} \left|\frac{\eta(\sigma)}
{\vartheta_1( \frac{\sigma}{4} +\frac{l}{4},\sigma)}\right|^2\right) .\qquad
\eea

\paragraph{Lattice 5: the $\Z_6$ orbifold.}

Lattice 5 has $\tau = e^{2\pi i/3}$, and by~(\ref{zmpara}) we may pick
arbitrary $\rho\in\H/(\PSL(2,\Z)$.  The untwisted sector partition
function is
$$
Z_u = \frac{1}{6}\left(Z(\tau = e^{2\pi i/3}, \rho) +
2\left|\frac{\eta(\sigma)}{\vartheta_2(\frac{1}{3},\sigma)}\right|^2 +
6\left|\frac{\eta(\sigma)}{\vartheta_1(\frac{1}{3},\sigma)}\right|^2 +
4\left|\frac{\eta(\sigma)}{\vartheta_2(\sigma)}\right|^2 \right) .
$$
Tori with $\tau = e^{2\pi i/3}$ have three fixed points under the
$\Z_6$ rotation symmetry, one corresponding to $\Z_2$, $\Z_3$, and
$\Z_6$ twists each.  The $\Z_6$ orbifold partition function therefore
is the sum of untwisted, $\Z_2$, $\Z_3$, and $\Z_6$ twisted sector
partition functions
$$\label{z6or}
Z_{\Z_6-{\rm orb}}(\tau = e^{2\pi i/3}, \rho) = 
Z_u + Z_{2t} + Z_{3t} + Z_{6t}\, .
$$
As before, the $\Z_2$ twisted sector partition function $Z_{2t}$ is
obtained from~(\ref{z2tiw}) by omitting the factor of four.  The
$\Z_3$ twisted sector partition function $Z_{3t}$ can be read off
from~(\ref{z3tiw}) by omitting the factor of three.  By~(\ref{dim})
the ground states in each of the twisted sectors ${\mathcal
H}_{\gamma}$, ${\mathcal H}_{\gamma^5}$ have dimensions $(h, \bar h) =
(5/72,5/72)$, and the $\Z_6$ twisted sector partition function is
\begin{eqnarray*}
Z_{6t} &=& \frac{1}{6}\left(
2\left|\frac{\eta(\sigma)}{\vartheta_3(\frac{1}{3},\sigma)}\right|^2 +
2\left|\frac{\eta(\sigma)}{\vartheta_4(\frac{1}{3},\sigma)}\right|^2 +
\left|\frac{\eta(\sigma)}{\vartheta_3(\sigma)}\right|^2 +
\left|\frac{\eta(\sigma)}{\vartheta_4(\sigma)}\right|^2+\right.
\\ && 
	\hphantom{\frac 16\Biggl(}\!
\left. +2(q\bar{q})^{-{1}/{18}} \sum_{l=-1}^{1}\sum_{i=1}^{4}
\left|\frac{\eta(\sigma)} {\vartheta_i(\frac{l}{3} +
\frac{\sigma}{3},\sigma)}\right|^2\right) .
\end{eqnarray*}
Altogether, we obtain
\begin{eqnarray}
Z_{\Z_6-{\rm orb}}(\tau\!=e^{2\pi i/3}\!, \rho) 
\!& = \!&\frac{1}{6}\!\left( Z(\tau=e^{2\pi i/3}, \rho) +
4\sum_{j=2}^{4}\left|\frac{\eta(\sigma)} 
{\vartheta_j(\sigma)}\right|^2+\right.
\nonumber\\&&
	\hphantom{\frac{1}{6}\!\Biggl( }\!
+ 2\sum_{i=1}^{4}\left|\frac{\eta(\sigma)}
{\vartheta_i(\frac{1}{3},\sigma)}\right|^2 +
4\left|\frac{\eta(\sigma)}{\vartheta_1(\frac{1}{3},\sigma)}\right|^2+
\label{lattice5}\\[3pt]&& 
	\hphantom{\frac{1}{6}\!\Biggl( }\!
\left. +\,(q\bar{q})^{-\frac{1}{18}}\! \sum_{l=-1}^{1}\!\left(
2\sum_{i=1}^{4}\! \left|\frac{\eta(\sigma)} {\vartheta_i(\frac{l}{3} {+}
\frac{\sigma}{3},\sigma)}\right|^2\!\! + 4\left|\frac{\eta(\sigma)}
{\vartheta_1(\frac{l}{3} {+} \frac{\sigma}{3},\sigma)}\right|^2\right)
\!\right) \!.
\nonumber
\end{eqnarray}

\subsection{Lattices 6 to 17: modding out by $R$ or 
$T_R^{(\prime)}$ reflection symmetries}

The reflection symmetry $R$ is a symmetry for every lattice with
$\tau_1 \in \{0, {1/2}\}$.  By inspection of the action on the
respective fundamental cell one easily checks that an exchange of
$R_1$ and $R_2$ is equivalent to a transformation of $\tau_2$; we
define
\begin{eqnarray*}
S, T\in \PSL(2,\Z):\quad\quad 
S:\zeta\mapsto -{1\over\zeta},\quad
T:\zeta\mapsto \zeta+1;\\
\Theta:= \left\{\begin{array}{ll}
S &\mbox{ if } \zeta_1=0,\\
TS\mathbb{T}^2S &\ds \mbox{ if } \zeta_1={1\over2}.
\end{array}\right.
\end{eqnarray*}
Then
\be\label{exchange}
R_1 \leftrightarrow R_2 \mbox{ is equivalent to }
\tau\mapsto \Theta\tau, \mbox{ where }
\left\{\begin{array}{l}
\Theta(i\tau_2)=\ds{i\over\tau_2}\,, 
\\[4pt]
\ds\Theta\left({1\over2}+i\tau_2\right)={1\over2}+{i\over4\tau_2}\,.
\end{array}\right.
\ee
We can therefore restrict ourselves to the discussion of the symmetry
$R_1$ in the following.  To extend $R_1$ to the charge
lattice~(\ref{latt}), the B-field $B$ must obey~(\ref{cont}) which is
true iff $\rho_1\in{1\over2}\Z$. Then by using~(\ref{char}) for the
$R_1$ action $|m_1,m_2,n_1,n_2\rangle\mapsto
|m_1^\prime,m_2^\prime,n_1^\prime,n_2^\prime\rangle$ we obtain
\be\label{ground1}
\begin{array}[b]{rclrcl}
m'_1 &=& -m_1\,,  &
n'_1 &=& -n_1 + 2\tau_1 n_2 + 2\rho_1 m_2 + 4\tau_1\rho_1 m_1\,, 
\\[3pt]
m'_2 &=& m_2 + 2\tau_1 m_1 \,,\qquad& n'_2 &=& n_2 + 2\rho_1 m_1\,,
\end{array}
\ee
and the invariant vectors $(p,\bar p)$ of the charge lattice
correspond to $|0,m_2,n_1,n_2\rangle$,
\be
\label{invariant}
\mpbar{p} = {1\over\sqrt{2\tau_2\rho_2} } \left( n_2\tau_2\pm
m_2\rho_2 \atop 0 \right) , \quad n_2,m_2\in\Z \mbox{ such that }
n_1=n_2\tau_1+ m_2\rho_1\in\Z\,.
\ee
The Hilbert space ground states $|m_1,m_2,n_1,n_2\rangle$ are pairwise
orthogonal, so the only states that give a contribution to
${\scriptstyle{R_1}}\square_{1}\limits$ are the ones that are built by
an action of creation operators on ground states corresponding to
vertex operators with $R_1$-invariant charge vectors
(\ref{invariant}).

Because~(\ref{invariant}) only depends on $\rho_1\mod\Z$ the same is
true for the resulting orbifold theory and we can pick
$\rho_1\in\{0,{1/2}\}$.  Note that in the case $\rho_1={1/2}$ the
B-field of our theory is effectively shifted by an integer form if we
apply $R_1$. This will be of some importance below.

To understand the action of the symmetry
$T_R^{(\prime)}=RT_{\delta^{(\prime)}}$ on the Hilbert space of a
toroidal conformal field theory observe that $T_{\delta^{(\prime)}}$
only acts on the ground state sectors and leaves the oscillator modes
invariant. On a state $|m_1,m_2,n_1,n_2\rangle$ corresponding to the
charge vector $(p,\bar p)(\lambda,\mu)$ the action of
$T_{R_1}^{(\prime)}$ is given by the action (\ref{ground1}) of $R_1$
combined with multiplication by $\exp[2\pi i (p,\bar
p)(\lambda,\mu)\cdot {1\over2}(p,\bar p)(2\delta^{(\prime)},0)]
=(-1)^{\langle\mu,2\delta^{(\prime)}\rangle}$, where we
used~(\ref{charges}).  It is therefore a priori clear that as for the
action of $R$ we need to restrict the possible B-field values to
$\rho_1\in\{0,{1/2}\}$ for consistency of the action of
$T_R^{(\prime)}$.  By~(\ref{group}), $T_R^{(\prime)}$ actions are only
needed in the case $\tau_1=0$. Using~(\ref{char}) one now checks that
only for $\rho_1=0$ the order of $T_R^{(\prime)}$ is two, whereas for
$\rho_1=1/2$ we find that $T_R^{(\prime)}$ generates a $\Z_4$ type
group.  The action of $g:=(T_R^{(\prime)})^2$ is given by
multiplication with $\pm1$ on the different Hilbert space sectors. To
mod out a toroidal theory $A$ by this $\Z_4$ then is equivalent to
performing a $\Z_2$ orbifold procedure on $A/\{1,g\}$. But $A/\{1,g\}$
is another toroidal theory, because both generic torus currents are
invariant under $g$ and give conserved currents in $A/\{1,g\}$ as
well. The $T_R^{(\prime)}$ action with $\rho_1=1/2$ hence need not be
considered separately.  For $\rho_1=0$ by~(\ref{exchange}) we now have
\be\label{texchange}
T_{R_1}^{(\prime)} \leftrightarrow T_{R_2}^{(\prime)} \mbox{ is equivalent to }
\tau\left(=i\tau_2\right)\mapsto \Theta\tau\left(={i\over\tau_2}\right).
\ee
Since by~(\ref{symmetries}) 
$\delta^{(\prime)}=\sqrt{{\rho_2/\tau_2}}({1/2\atop\ast})$,
if $|m_1,m_2,n_1,n_2\rangle$ is $R_1$-invariant, then
by~(\ref{invariant}) $m_1=0, n_1=n_2\tau_1+m_2\rho_1$, and
$T_R^{(\prime)}$ acts by
\be\label{taction}
T_R^{(\prime)}:\quad 
|m_1,m_2,n_1,n_2\rangle\mapsto (-1)^{n_2}|m_1,m_2,n_1,n_2\rangle.
\ee
Below we will construct the families of conformal field theories
obtained by the orbifold procedure with a group $G$ which corresponds
to one of the lattices 6 to 17.  This will yield irreducible
components $\mathcal{C}^{(\tau_1,\rho_1)}_{G-{\rm orb}}$ of the moduli
space $\mathcal{C}^2$ with $\tau_1,\rho_1\in\{0,{1/2}\}$. In some
cases discrete torsion gives additional degrees of freedom, increasing
the number of irreducible components to
$\mathcal{C}^{(\tau_1,\rho_1)}_{G^\pm-{\rm orb}}$ or even
$\mathcal{C}^{(\tau_1,\rho_1)}_{G^{\pm\pm}-{\rm orb}}$.  The Teichm\u
ller space of each such irreducible component is $(\R^+)^k$, where
$k=1$ if $\tau_2$ must be fixed for the particular lattice, too, and
$k=2$ otherwise.  To find the correct parameter spaces, we must
determine the subgroup $\mathcal{P}$ of $\PSL(2,\Z)^2\rtimes\Z_2^2$
in~(\ref{modspace}) which maps the respective Teichm\u ller space
$(\R^+)^k$ onto itself.  Then we must discuss which elements of
$\mathcal{P}$ map equivalent orbifold theories onto each other.

Restrict $\mathcal{P}$ to one of the factors $\R^+\subset\H$ of the
Teichm\u ller space $(\R^+)^k$, specified by $\zeta_1=0$ or
$\zeta_1={1/2}$.  We claim that
\be\label{tduality}
\mathcal{P}\cap \PSL(2,\Z)=\{1,\Theta\}\, .
\ee
As stated in~(\ref{exchange}), $\Theta$ acts on
$I^0:=\{\zeta\in\H\mid\zeta_1=0\}$ by $\zeta_2\mapsto{1/\zeta_2}$
and on $I^+:=\{\zeta\in\H\mid\zeta_1={1/2}\}$ by
$\zeta_2\mapsto{1/4\zeta_2}$. Now $I^0=J^0\cup \Theta J^0$, where
$J^0:=\{\zeta\in I^0\mid \zeta_2\geq1\}$. Because $J^0$ does not
contain any two points identified by M\o bius transformations, the
assertion follows for the case $\zeta_1=0$. For $\zeta_1={1/2}$
observe that $I^+=(J^+ \cup TST J^1)\cup \Theta(J^+ \cup TST J^1)$,
where $J^+ := \{\zeta\in I^+\mid \zeta_2\geq {\sqrt3/2} \}$ and
$J^1 := \{\zeta \in \H\mid \|\zeta\|=1, \zeta_1\in [-{1/2},0] \}$.
Because no two points in $J^+\cup J^1$ are related by M\o bius
transformations, the assertion follows.  For the respective factor of
the Teichm\u ller space under discussion, $\Theta$ will be called
T-duality.

By our convention to fix $\tau_1,\rho_1\in\{0,{1/2}\}$ it is clear
that target space orientation change $V:(\tau,\rho)\mapsto
(-\bar\tau,-\bar\rho)$ in~(\ref{UandV}) can only be contained in
$\mathcal{P}$ if $\tau_1=\rho_1=0$, in which case it acts
trivially. Mirror symmetry $U:(\tau,\rho)\mapsto (\rho,\tau)$ is
contained iff $\tau_1=\rho_1$ and $\tau_2$ is not fixed.  Inspection
of the charge lattice~(\ref{latt}) and the action (\ref{ground1}) of
$R_1$ shows that mirror symmetry commutes with $R_1,R_2$ on toroidal
conformal field theories.  But a priori it is not clear whether it
indeed commutes with the action of each of the symmetry groups
corresponding to lattices 6 to 17. Therefore, a case by case study is
necessary to decide which of $\Theta,U$ map a $G$ orbifold onto an
equivalent one and thus determine all the parameter spaces
$\mathcal{C}_{G^\bullet-{\rm orb}}^{(\bullet)}$.  We will also see
that not all of the lattices yield different components of the moduli
space $\mathcal{C}^2$.

\subsection{Lattices 6 and 7: the $\Z_2(R)$ reflection orbifold }

Lattices 6 ($\tau_1 = 0$) and 7 ($\tau_1={1/2}$) have reflection
symmetry $\Z_2(R)$.  For $\tau_1 = \rho_1 = 0$ the torus theory is the
product of two $c=1$ theories corresponding to compactification on a
circle each (see~(\ref{circle2})). The symmetry $R_1$ by~(\ref{refl})
leaves the first factor invariant and acts on the second as ordinary
$\Z_2$ orbifold.  Therefore the resulting partition function is the
product of circle and circle orbifold partition function; namely,
setting $r:= \sqrt{\rho_2/\tau_2}$, $r^\prime:=\sqrt{\tau_2\rho_2}$ as
in~(\ref{circle2}),
\be\label{lattice6}
Z_{R_1-{\rm orb}}(0,\tau_2,0,\rho_2) = 
Z^{c=1}(r) Z^{c=1}_{\rm orb}(r^\prime) \,,
\ee
where $Z^{c=1}$ and $Z^{c=1}_{\rm orb}$ are given in
(\ref{circlepart}) and~(\ref{c1orb}), respectively. If we mod out by
$R_2$ instead of $R_1$, by~(\ref{exchange}) we use
$\tau_2\mapsto{1/\tau_2}$, i.e.\ the radii $r$ and $r^\prime$ are
interchanged in~(\ref{lattice6}). Application of T-duality to both
$\tau$ and $\rho$ simultaneously, which will be denoted by
$$
\mathcal{S}:\; (\tau,\rho)\mapsto \left(-{1\over\tau}, -{1\over\rho}\right)
$$ 
and called simultaneous T-duality in the following, amounts to
$r\mapsto{1/ r}$, $r^\prime\mapsto {1/ r^\prime}$ in both cases,
leaving~(\ref{lattice6}) invariant. Mirror symmetry
$\tau_2\leftrightarrow\rho_2$ acts by $r\mapsto{1/r}$,
$r^\prime\mapsto r^\prime$, which (\ref{lattice6}) is invariant under,
too.

By~(\ref{gorb}) the general reflection orbifold partition function can
be written as
\be
\label{8orb}
Z_{R_1-{\rm orb}} = \frac{1}{2}\left( {\scriptstyle{1}}\square_{1} +
{\scriptstyle{R_1}}\square_{1} + {\scriptstyle{1}}\square_{R_1} +
{\scriptstyle{R_1}}\square_{R_1} \right) .
\ee 
As explained in our general discussion for lattices 6 to 17, the
second term in~(\ref{8orb}) gets contributions only from the states
built by an action of creation operators on such Hilbert space ground
states that are invariant under the action of $R_1$. The corresponding
charge vectors are given in ~(\ref{invariant}), namely for lattice 6
with $\tau_1=0, \rho_1={1/2}$ we obtain
\be
\label{spinv}
\mpbar p = {1\over\sqrt{2\tau_2\rho_2}} \left( n\tau_2 \pm
2m\rho_2\atop0\right) = \left( {n\over r}\pm mr\atop0 \right), \qquad
m,n\in\Z, r := \sqrt{\frac{2\rho_2}{\tau_2}}\,.
\ee
Therefore for the 
untwisted sector partition function 
\begin{eqnarray*}
Z_u \left(0,\tau_2,\frac{1}{2},\rho_2\right) &=& \frac{1}{2} \left( Z
+ \left|\frac{\vartheta_3\vartheta_4}{\eta^4}\right| \sum_{m,n}
q^{\frac{1}{2}\left(\frac{n}{r} + m r \right)^2}
\bar{q}^{\frac{1}{2}\left(\frac{n}{r} - m r\right)^2}\right) .
\end{eqnarray*}
The twisted sector partition 
function can be calculated by modular transformations, and  
the  complete reflection orbifold partition function is
\bea
\label{lattice66}
Z_{R_1-{\rm orb}}(0,\tau_2,\frac{1}{2},\rho_2) &=& \frac{1}{2} \Bigg(
Z + \left|\frac{\vartheta_3\vartheta_4}{\eta^4}\right| \sum_{m,n}
q^{\frac{1}{2}\left(\frac{n}{r} + m r \right)^2}
\bar{q}^{\frac{1}{2}\left(\frac{n}{r} - m r\right)^2}+
\nonumber\\ && 
	\hphantom{\frac12\Biggl(}\!
+\left|\frac{\vartheta_3\vartheta_2}{2\eta^4}\right| \sum_{m,n}
q^{\frac{1}{8}\left(\frac{n}{r} + m r\right)^2}
\bar{q}^{\frac{1}{8}\left(\frac{n}{r} - m r \right)^2}+
\nonumber\\ && 
	\hphantom{\frac12\Biggl(}\!
\left.+\left|\frac{\vartheta_4\vartheta_2}{2\eta^4}\right|
 \sum_{m,n}(-1)^{mn} q^{\frac{1}{8}\left(\frac{n}{r} + m r\right)^2}
 \bar{q}^{\frac{1}{8}\left(\frac{n}{r} - m r\right)^2}\right) ,\qquad
\eea
with $r=\sqrt{2\rho_2/\tau_2}$, and for $R_2$ instead of $R_1$ with
$r=\sqrt{2\tau_2\rho_2}$ by~(\ref{exchange}). Simultaneous T-duality 
$\mathcal{S}$ amounts to 
$r\mapsto{1/ r}$ in both cases.
This obviously leaves~(\ref{lattice66}) invariant.

Mirror symmetry $U:(\tau,\rho)\mapsto (\rho,\tau)$ commutes with the
$R_1, R_2$ actions on a toroidal theory, so 
\be
\label{lattice7}
Z_{R-{\rm orb}}\left({1\over2},\tau_2,0,\rho_2\right) 
= Z_{R-{\rm orb}}\left(0,\rho_2,{1\over2},\tau_2\right).
\ee
Hence the partition function $Z_{R-{\rm orb}}({1/2},\tau_2,0,\rho_2)$
for lattice 7 with $\rho_1=0$ is given by~(\ref{lattice66}), but now
with $r=\sqrt{\rho_2/2\tau_2}$ for $R_1$, $r=\sqrt{2\tau_2\rho_2}$ for
$R_2$. Again, $\mathcal{S}$ acts by $r\mapsto{1/ r}$ in both cases
and leaves the partition function invariant.

The case $\tau_1=\rho_1 = 1/2$ (lattice 7) is more subtle.  The charge
lattice~(\ref{char}) of the toroidal theory is generated by the four
vectors
$$
v_{\delta,\epsilon}:= {1\over2\sqrt{2\tau_2\rho_2}} \left(
\pmatrix{\tau_2+\delta\rho_2\cr 
	\epsilon\,(1/2-2\delta\tau_2\rho_2)} ,
\pmatrix{\tau_2-\delta\rho_2\cr
	\epsilon\,(1/2+2\delta\tau_2\rho_2)} 
\right), \qquad \delta,\epsilon\in\{\pm1\}\,,
$$
which are pairwise interchanged by $R_1$ 
($v_{\delta,1}\leftrightarrow v_{\delta,-1}$).  Denote the corresponding
vertex operators by $V(\pm v_{\delta,\epsilon})$. The $R_1$ invariant
part of the charge lattice by~(\ref{invariant}) is given by
\bea
\label{hhinv}
&\displaystyle 
\mpbar{p}  = {1\over\sqrt{2\tau_2\rho_2} }
\left( n_2\tau_2\pm m_2\rho_2 \atop 0 \right) 
= \sqrt{2}\left( {n\over r}\pm mr \atop 0 \right) , 
\qquad\qquad\qquad\qquad
\\&\displaystyle \qquad\qquad\qquad\qquad
n_2=2n,\quad m_2=2m,\quad n_1=n+m\in\Z,\quad
r = \sqrt{\rho_2\over \tau_2}\,.
\nonumber
\eea
Because $\langle v_{\delta,\epsilon}, v_{\delta, -\epsilon} \rangle=1$,
the  vertex operators corresponding to generators of the invariant part of the
charge lattice are obtained from operator product
expansions
$$
\left( V(v_{\delta,1})+V(-v_{\delta,1})\right)
\times \left( V(v_{\delta,-1})-V(-v_{\delta,-1})\right).
$$
Since this is a product between an $R_1$ even and an $R_1$ odd
operator, the resulting vertex operators are $R_1$ odd. It follows
that $R_1$ acts on ground states corresponding to invariant charge
vectors~(\ref{hhinv}) by
$|m_1,m_2,n_1,n_2\rangle\mapsto(-1)^{n_2}|m_1,m_2,n_1,n_2\rangle$.
Thus for the untwisted sector partition function we find
$$
Z_u \left(\frac{1}{2},\tau_2,\frac{1}{2},\rho_2\right) =\frac{1}{2}
\left( Z + \left|\frac{\vartheta_3\vartheta_4}{\eta^4}\right|\left(
\sum_{m,n\in\Z} - \sum_{m,n\in{\Z + 1/2}} \right) q^{\left(\frac{n}{r}
+ m r \right)^2} \bar{q}^{\left(\frac{n}{r} - m r\right)^2}\right),
$$
with $r=\sqrt{\rho_2/\tau_2}$.  We remark that although not stated
explicitly above, one may check that in none of the other cases of
$R_1$ actions such additional signs on Hilbert space ground states
occur. Here, they are due to the fact that the action of $R_1$
effectively shifts the B-field by an integer form, as was already
mentioned above.  In the discussion of the bicritical
point~(\ref{t_z3}) we will point out a very natural confirmation of
the above result.  By applying modular transformations to
$\smash{{\scriptstyle{R_1}}\square_{1}\limits\,}$, we find
\bea
Z_{R_1-{\rm orb}}\!\left(\frac{1}{2},\tau_2,\frac{1}{2},
\rho_2\right)\! &=& \frac{1}{2} \!\left( Z +
\left|\frac{\vartheta_3\vartheta_4}{\eta^4}\right|
\left(\sum_{m,n\in\Z}\! -\!\! \sum_{m,n\in{\Z + 1/2}} \right)\!
q^{\left(\frac{n}{r} + m r \right)^2}\! \bar{q}^{\left(\frac{n}{r} - m
r\right)^2}\!+\right.
\label{lattice77}\\ & & 
	\hphantom{\frac{1}{2} \Biggl(}\!
+\left|\frac{\vartheta_3\vartheta_2}{4\eta^4}\right|
\!\sum_{m,n\in\Z}\!\left(1 \!-\! (-1)^{n + m}\right)
q^{\frac{1}{16}\left(\frac{n}{r} + m r\right)^2}
\bar{q}^{\frac{1}{16}\left(\frac{n}{r} - m r \right)^2}\!+
\nonumber\\ && 
	\hphantom{\frac{1}{2} \Biggl(}\!
+\!\left.\left|\frac{\vartheta_4\vartheta_2}{4\eta^4}\right|
\!\sum_{m,n\in\Z}\!\left(1 \!-\! (-1)^{n + m}\right)(i)^{nm}
  q^{\frac{1}{16}\left(\frac{n}{r}  + m r\right)^2}
\!\bar{q}^{\frac{1}{16}\left(\frac{n}{r}  - m r\right)^2}\!\right)\! ,
\nonumber
\eea
where $r = \sqrt{\rho_2/\tau_2}$ for the reflection $R_1$, and
therefore $r= 2\sqrt{\tau_2\rho_2}$ for the reflection $R_2$
by~(\ref{exchange}).  To apply simultaneous T-duality $\mathcal{S}$
amounts to $r\mapsto{1/r}$, yielding~(\ref{lattice77}) invariant.
Invariance under mirror symmetry $\tau_2\leftrightarrow\rho_2$ is also
obvious.

By our general discussion for lattices 6 to 17, to find the correct
parameter space for the irreducible components of $\mathcal{C}^2$
obtained by $\Z_2(R)$ orbifolding, the Teichm\u ller spaces are
constructed by considering $R=R_1$ only. T-duality applied to $\tau$
alone, which by~(\ref{exchange}) is equivalent to $R_1\leftrightarrow
R_2$, does not generically map onto an isomorphic theory. Our
calculations above show that simultaneous T-duality $\mathcal{S}$
actually identifies isomorphic theories (see~(\ref{lattice6}),
(\ref{lattice66}) and~(\ref{lattice77})) as well as mirror symmetry.
In particular, lattice 6 ($\tau_1=0$) with $\rho_1={1/2}$ and
lattice 7 ($\tau_1={1/2}$) with $\rho_1=0$ correspond to families
of isomorphic orbifold conformal field theories.

Summarizing, we have constructed the following three irreducible
components of the moduli space:
\begin{eqnarray*}
\mathcal{C}_{\Z_2(R)-{\rm orb}}^{(0,0)} 
&\cong& \left(\R^+\right)^2/\{U,\mathcal{S}\}\,,\qquad
\quad\quad\mathcal{C}_{\Z_2(R)-{\rm orb}}^{({1\over2},{1\over2})} 
\cong \left(\R^+\right)^2/\{U,\mathcal{S}\}
\\
\mathcal{C}_{\Z_2(R)-{\rm orb}}^{(0,{1\over2})} 
&=& \mathcal{C}_{\Z_2(R)-{\rm orb}}^{({1\over2},0)}
\cong \left( \R^+ \right)^2/\mathcal{S}\,. 
\end{eqnarray*}

\subsection{Lattices 8 and 9: the $D_2$ orbifold}

Lattices 8 $(\tau_1 = 0)$ and 9 $(\tau_1 = 1/2)$ have a $D_2 = \{1,
A(\pi), R_1, R_2\}$ symmetry.  By~(\ref{gorb}) for both
$\tau_1,\rho_1\in\{0,{1/2}\}$ the $D_2$ orbifold partition
function is
\bea
\label{d2pre}
Z_{D_2-{\rm orb}}(\tau_1,\tau_2,\rho_1,\rho_2) & = &
\frac{1}{4}\sum_{g,h\in D_2} {\scriptstyle{g}}\square_{h}
\nonumber\\
& = & \frac{1}{4} \left( 2 Z_{\Z_2-{\rm orb}} + 2Z_{R_1-{\rm orb}} +
2Z_{R_2-{\rm orb}} - 2Z +
\vphantom{{\scriptstyle{R_1}}\square_{A(\pi)}}\right.
\\& & 
\hphantom{\frac{1}{4} \Biggl(}\!
\left.  + {\scriptstyle{R_1}}\square_{A(\pi)}
+{\scriptstyle{A(\pi)}}\square_{R_1} +
{\scriptstyle{R_2}}\square_{R_1} + {\scriptstyle{R_2}}\square_{A(\pi)}
+ {\scriptstyle{A(\pi)}}\square_{R_2} +
{\scriptstyle{R_1}}\square_{R_2} \right) ,
\nonumber
\eea
where we have subtracted $Z= {\scriptstyle{1}}\square_{1}\limits$ from
the second and third term in the second line to avoid overcounting the
contribution of the identity element which appears in each reflection
group.  Observe that by~(\ref{exchange}) separate T-duality
(\ref{tduality}) on $\tau$, or by mirror symmetry equivalently on
$\rho$, interchanges $\Z_2(R_1)$ and $\Z_2(R_2)$. Therefore it maps
isomorphic $D_2$ orbifold conformal field theories onto each other.

The terms in the third line of~(\ref{d2pre}) form a modular orbit. To
determine them we compute
${\scriptstyle{R_1}}\square_{A(\pi)}\limits$.  Denote by ${\mathcal
H}_{A(\pi)}$ the twisted sector Hilbert space of the ordinary $\Z_2$
orbifold which by~(\ref{zmmod}) corresponds to fields $\varphi$ with
half integer modes and $\varphi(z=0) = q_j$, $j\in\{1,2,3,4\}$, a
$\Z_2$ fixed point on $\mathbb{T}^2$. Assume that $k$ of the four
corresponding $\Z_2$ twisted ground states are eigenstates of
$R_1$. There eigenvalues must agree and be $\pm1$ in order for the
$\Z_2$ action on the twisted sector to be well defined.  Since
by~(\ref{dim}) the twisted ground states have dimensions $(h,\bar h) =
(1/8,1/8)$, we find
\bea
\label{tos}
{\scriptstyle{R_1}}\square_{A(\pi)} &=& \tr_{{\mathcal H}_{A(\pi)}}
R_1 q^{L_0 -\frac{c}{24}}\bar{q}^{\bar L_0 -\frac{c}{24}}
\nonumber\\[-4pt]
&=& \pm k\cdot (q\bar q)^{-{1}/{12}}\frac{(q\bar q)^{{1}/{8}}}
{\prod_{n=1}^{\infty}(1 - q^{n-1/2})(1 - \bar{q}^{n-1/2}) (1 +
q^{n-1/2})(1 + \bar{q}^{n-1/2})}
\nonumber\\
& = & \pm k\left|\frac{\eta^2}{\vartheta_3\vartheta_4}\right| =
\pm{k\over2}\left|\frac{\vartheta_2}{\eta}\right|.
\eea
All in all by modular transformations the third line in~(\ref{d2pre})
is equal to 
$\pm 2k Z_{\rm Ising}$, where
\be\label{Ising}
Z_{\rm Ising}=  \frac{1}{2}\left(
                          \left|\frac{\vartheta_2}{\eta}\right| 
                         + 
              \left|\frac{\vartheta_4}{\eta}\right|
                         +
 \left|\frac{\vartheta_3}{\eta}\right| \right)
\ee
and $k\in\{0,2,4\}$,
because $R_1$ must map twisted ground states onto twisted ground states. 
To determine the correct factor $k$ we first
note that in case $\tau_1=\rho_1=0$ the original toroidal theory
decomposes into a tensor product of two $c=1$ theories. The action
of $D_2$ respects the product structure, hence
\be
\label{lattice80}
Z_{D_2^+-{\rm orb}}(0,\tau_2,0,\rho_2) = Z_{\rm
orb}^{c=1}(\sqrt{\tau_2\rho_2})Z_{\rm
orb}^{c=1}(\sqrt{\rho_2/\tau_2})\,,
\ee
where $Z_{\rm orb}^{c=1}$ was given in~(\ref{c1orb}).  One now checks
that in this case $k=4$, in agreement with the geometric observation
that all the four $\Z_2$ fixed points on $\mathbb{T}^2$ are invariant
under the $R$ actions. For $\tau_1=1/2, \rho_1=0$ one can argue that
only two of the four fixed points are invariant, thus $k=2$. If
$\rho_1=1/2$, this geometric argument breaks down since, as noted in
our general discussion for lattices 6 to 17, in this case the
symmetries $R_1,R_2$ effectively shift the B-field by an integer
form. The correct factor for $\tau_1=0,\rho_1=1/2$ is $k=2$, as well.
This follows from the construction of the $D_4$ orbifold conformal
field theory (lattice 15), where we will see that the $D_2$ orbifold
at $\tau_1=0,\rho_1=1/2$ must always contain an even number of fields
with dimensions $h=\bar h=1/16$.  For $\tau_1=\rho_1=1/2$ we find
$k=0$. This follows from the fact that
${\scriptstyle{1}}\square_{R_1}\limits$ by (\ref{lattice77})
generically does not get any contributions from fields with dimensions
$h=\bar h=1/16$. Hence $\smash{{\scriptstyle{A(\pi)}}\square_{R_1}\limits
=\pm{k\over2}\left|{\vartheta_3\over\eta}\right|}$ cannot give such
contributions either.  In summary,
\bea
\label{lattice8}  
Z_{D_2^\pm-{\rm orb}}(\tau_1,\tau_2,\rho_1,\rho_2) &=& \frac{1}{2}
(Z_{\Z_2-{\rm orb}} + Z_{R_1-{\rm orb}} + Z_{R_2-{\rm orb}} \pm
kZ_{\rm Ising} - Z )\,,
\nonumber\\
k&=&4(1-\tau_1-\rho_1)\,,
\eea        
where $Z_{\Z_2-{\rm orb}}$ is given in ~(\ref{lattice2}), and
$Z_{R-{\rm orb}}$ is given in~(\ref{lattice6}),~(\ref{lattice66}) or
(\ref{lattice77}), respectively.  In particular, for this orbifold
construction discrete torsion has a nontrivial effect, and we can
produce two non-equivalent theories corresponding to lattice 8 and
each possible value of $\rho_1$.  We stress that we have been
discussing a perhaps counterintuitive effect of ``turning on the
B-field'': the action of $R_1,R_2$ on twisted ground states depends
severely on the value of $\rho_1$. In particular, they must not be
interpreted from a purely geometric point of view.
\pagebreak[3]

Because $\Z_2(R)$ orbifold conformal field theories as well as the
formula for $k$ in~(\ref{lattice8}) are invariant under mirror
symmetry, the same is true for $D_2$ orbifolds. Hence we have
constructed five irreducible components of $\mathcal{C}^2$,
$$
\mathcal{C}_{D_2^\pm-{\rm orb}}^{(0,0)} 
\cong \left(\R^+/\Theta\right)^2/U, \quad
\mathcal{C}_{D_2^\pm-{\rm orb}}^{(0,{1\over2})} 
\cong \mathcal{C}_{D_2^\pm-{\rm orb}}^{({1\over2},0)}
\cong \left(\R^+/\Theta\right)^2, \quad
\mathcal{C}_{D_2-{\rm orb}}^{({1\over2},{1\over2})} 
\cong \left(\R^+/\Theta\right)^2/U.
$$

\subsection{Lattice 10: the $\Z_2(T_R)$ reflection plus shift orbifold} 

Lattice 10 $(\tau_1 = 0)$ has reflection plus shift symmetry
$\Z_2(T_{R_1}) = \{1, T_{R_1}=$ \linebreak $ R_1 e^{2\pi i p \cdot
{\delta_1/\sqrt 2}} \}$, where $\delta_1 = \sqrt{\rho_2/\tau_2}(1/2,
0)$.  From our general discussion on the $T_R^{(\prime)}$ action for
lattices 6 to 17 we know that we only have to consider the case
$\rho_1=0$.  By~(\ref{gorb}) the general reflection plus shift
orbifold partition function is
\be
\label{gtr}
Z_{T_{R_1}-{\rm orb}} = \frac{1}{2} \left(
{\scriptstyle{1}}\square_{1} + {\scriptstyle{T_{R_1}}}\square_{1} +
{\scriptstyle{1}}\square_{T_{R_1}} +
{\scriptstyle{T_{R_1}}}\square_{T_{R_1}} \right).
\ee
The torus theory is a tensor product of two $c=1$ circle
theories~(\ref{circle2}): ${\scriptstyle{1}}\square_{1}\limits$ $=
{\scriptstyle{1}}\square_{1}\limits(\varphi_1)
{\scriptstyle{1}}\square_{1}\limits(\varphi_2)$.  The $\Z_2(T_{R_1})$
action respects the product structure, therefore we have
${\scriptstyle{T_{R_1}}}\square_{1}\limits =
{\scriptstyle{t_{\epsilon}}}\square_{1}\limits(\varphi_1)
{\scriptstyle{(-1)}}\square_{1}\limits(\varphi_2)$ with $t_{\epsilon}
= e^{2\pi i p\cdot{\epsilon\over\sqrt2}},
\epsilon=\half\sqrt{\rho_2/\tau_2}$.  From the circle orbifold theory
(\ref{c1orb}) one has
${\scriptstyle{(-1)}}\square_{1}\limits(\varphi_2) =
2\left|{\eta}/{\vartheta_2}\right|$.  As explained in our general
discussion for lattices 6 to 17, the translation symmetry
$t_{\epsilon}$ does not affect oscillator modes. By~(\ref{invariant})
it acts on the Hilbert space ground states $|0,m,0,n\rangle$ of the
circle theory ${\scriptstyle{1}}\square_{1}\limits(\varphi_1)$ via
multiplication with $(-1)^n$. So using~(\ref{circlepart})
and~(\ref{invariant}) with $r:=\sqrt{\rho_2/\tau_2}$ we find
$$
{\scriptstyle{t_{\epsilon}}}\square_{1}(\varphi_1) =
\frac{1}{\eta\bar{\eta}} \sum_{m, n} (-1)^n q^{\frac{1}{4}({n\over r}
+ mr)^2} \bar{q}^{\frac{1}{4}({n\over r} - mr)^2}\,.
$$
The remaining boxes in~(\ref{gtr}) are obtained by modular
transformations. Thus the complete partition function is
\bea
Z_{T_{R_1}-{\rm orb}}(0,\tau_2, 0,\rho_2) &=& \frac{1}{2}\left (Z + 
\left|\frac{\vartheta_3\vartheta_4}{\eta^4}\right| \sum_{m,n}(-1)^n
q^{\frac{1}{4}\left(\frac{n}{r} + mr\right)^2}
\bar{q}^{\frac{1}{4}\left(\frac{n}{r} - mr\right)^2}+\right.
\nonumber\\&& 
	\hphantom{\frac{1}{2}\Bigg (}\!
+\left|\frac{\vartheta_3\vartheta_2}{\eta^4}\right| \sum_{n\in
\Z,m\in{\Z+1/2}} q^{\frac{1}{4}\left(\frac{n}{r} + mr\right)^2}
\bar{q}^{\frac{1}{4}\left(\frac{n}{r} - mr\right)^2}+
\nonumber\\&&
	\hphantom{\frac{1}{2}\Bigg (}\!
+\left.\left|\frac{\vartheta_4\vartheta_2}{\eta^4}\right| \sum_{n\in
\Z,m\in{\Z+1/2}}(-1)^n q^{\frac{1}{4}\left(\frac{n}{r} + mr\right)^2}
\bar{q}^{\frac{1}{4}\left(\frac{n}{r} - mr\right)^2}\right),\qquad
\label{lattice10}
\eea
where $r:= \sqrt{\rho_2/\tau_2}$. If we mod out by $\Z_2(T_{R_2})$
instead of $\Z_2(T_{R_1})$ by~(\ref{texchange}) we have to set
$r:=\sqrt{\tau_2\rho_2}$.  Simultaneous T-duality $\mathcal{S}$
amounts to $r\mapsto{1/r}$ which does not leave~(\ref{lattice10})
invariant. But for $\Z_2(T_{R_1})$ orbifolds, $\mathcal{S}$ combined
with mirror symmetry maps onto an isomorphic theory, whereas for
$\Z_2(T_{R_2})$ orbifolds mirror symmetry maps onto isomorphic
theories.  Therefore the $\Z_2(T_R)$ orbifold conformal field theories
form a family
$$
\mathcal{C}_{\Z_2(T_R)-{\rm orb}}\cong \left(\R^+\right)^2/U\,.
$$

\subsection{Lattice 11: the $D_2(T_R)$ orbifold}

Lattice 11 $(\tau_1 = 0)$ has a $D_2(T_R) = \{1, A(\pi),T_{R_1},
\widehat{T}_{R_2}\}$ symmetry as defined in~(\ref{symmetries}), and we
can generally set $\rho_1=0$.  By~(\ref{gorb}) the partition function
has the form
\begin{eqnarray}
Z_{D_2(T_R)-{\rm orb}}(0,\tau_2, 0,\rho_2) & = & \frac{1}{4} \Biggl( 2
Z_{\Z_2-{\rm orb}} + 2 Z_{T_{R_1}-{\rm orb}} +2
Z_{\widehat{T}_{R_2}-{\rm orb}} - 2Z + 
\label{l11}\\&&
	\hphantom{\frac14\Bigl(}
+ {\scriptstyle{T_{R_1}\!\!}}\square_{A(\pi)}
+ {\scriptstyle{{\widehat{T}_{R_2}\!\!}}}\square_{A(\pi)}    
+ {\scriptstyle{A(\pi)}}\square_{T_{R_1}}
+ {\scriptstyle{A(\pi)}}\square_{{\widehat{T}_{R_2}}}
+ {\scriptstyle{T_{R_1}}}\square_{{\widehat{T}_{R_2}}} 
+ {\scriptstyle{\widehat{T}_{R_2}\!}}\square_{T_{R_1}} \,
\Biggl).
\nonumber
\end{eqnarray}
The terms in the second line can be computed by a similar argument as
those in the third line of~(\ref{d2pre}). Only here none of the four
ordinary $\Z_2$ fixed points is invariant under $T_{R_1}$ or
$\widehat{T}_{R_2}$, so the first two boxes vanish. The others are
obtained by modular transformations from these and therefore vanish as
well. In particular, in this case discrete torsion has no effect on
the partition function.

The original toroidal theory decomposes into the tensor product of two
$c=1$ theories~(\ref{circle2}). By~(\ref{symmetries})
$\widehat{T}_{R_2}$ leaves the second factor invariant. Since on the
Hilbert space ground states $|0,m,0,n\rangle$ of the first factor
$\widehat{T}_{R_2}|0,m,0,n\rangle = \pm|0,-m,0,-n\rangle= \pm
R_2|0,m,0,n\rangle$,
$\smash{{\scriptstyle{\widehat{T}_{R_2}}}\square_{1}\limits
={\scriptstyle{R_2}}\square_{1}\limits}$ and therefore we have
$Z_{\widehat{T}_{R_2}-{\rm orb}} = Z_{R_2-{\rm orb}}$.  All in all
\be
\label{lattice11}  
Z_{D_2(T_R)-{\rm orb}}(0,\tau_2, 0,\rho_2) = \frac{1}{2}(Z_{\Z_2-{\rm
orb}} + Z_{R_2-{\rm orb}} + Z_{T_{R_1}-{\rm orb}} - Z ) \,,
\ee
where $Z_{\Z_2-{\rm orb}}, Z_{R_2-{\rm orb}}, Z_{T_{R_1}-{\rm orb}}$
are given in (\ref{lattice2}),~(\ref{lattice6}), and
~(\ref{lattice10}), respectively.

By the discussion of $\Z_2(T_R)$ orbifold conformal field theories
(lattice 10), only combined $\mathcal{S}$ with mirror symmetry leaves
$Z_{T_{R_1}-{\rm orb}}$ invariant. This also maps isomorphic $R$
orbifolds onto each other (lattice 6), so the $D_2(T_R)$ orbifold
conformal field theories form a family
$$
\mathcal{C}_{D_2(T_R)-{\rm orb}}
\cong \left(\R^+\right)^2/U\mathcal{S}.
$$

\subsection{Lattice 12: the $D_2(\mathbb{T}^\prime_R)$ orbifold}

Lattice 12 $(\tau_1 = 0)$ has a $D_2(\mathbb{T}^\prime_R) = \{1,
A(\pi),\mathbb{T}^\prime_{R_1}, \mathbb{T}^\prime_{R_2}\}$ symmetry as
defined in~(\ref{symmetries}), and we may set $\rho_1=0$.  The
calculation of the partition functions is analogous to that for
lattice 11, where in (\ref{l11}) we now replace $T_{R_1}$ by
$\mathbb{T}^\prime_{R_1}$ and $\widehat{T}_{R_2}$ by
$\mathbb{T}^\prime_{R_2}$. Again, none of the ordinary $\Z_2$ fixed
points is invariant under a symmetry ${T}_{R}^\prime$. So the second
line in~(\ref{l11})$^\prime$ vanishes, too, and discrete torsion has
no effect.

For $\tau_1 =\rho_1 = 0$ analogously to $Z_{\widehat{T}_{R_2}-{\rm
orb}} =Z_{R_2-{\rm orb}}$ in the partition function for lattice 11 we
now find $Z_{\mathbb{T}^\prime_{R}-{\rm orb}} =Z_{T_{R}-{\rm orb}}$.
So we have
\be
\label{lattice12}
Z_{D_2(\mathbb{T}^\prime_R)-{\rm orb}}(0,\tau_2, 0,\rho_2) =
\frac{1}{2}(Z_{\Z_2-{\rm orb}} + Z_{T_{R_1}-{\rm orb}} +
Z_{T_{R_2}-{\rm orb}} - Z )\,,
\ee
where $Z_{\Z_2-{\rm orb}}$ and $Z_{T_{R}-{\rm orb}}$ are given in
$(\ref{lattice2})$ and $(\ref{lattice10})$, respectively.  Since
T-duality~(\ref{texchange}) applied to $\tau$ interchanges
$\Z_2(T_{R_1})$ and $\Z_2(T_{R_2})$, but neither simultaneous
T-duality $\mathcal{S}$ nor mirror symmetry leaves invariant both of
them, $D_2(T_R^\prime)$ orbifold conformal field theories form a
family
$$
\mathcal{C}_{D_2(\mathbb{T}^\prime_R)-{\rm orb}}
\cong \left(\R^+/\Theta\right)\times\R^+\,.
$$

\subsection{Lattice 13: the $D_3(R_1)$ orbifold}

Lattice 13 $(\tau = e^{2\pi i/3})$ has a
$D_3(R_1) = \Z_3\cup\{ R_1, A(2\pi/3)R_1, A(4\pi/3)R_1\}$ symmetry.
By our general discussion for lattices 6 to 17 and since for lattice
13 the value of 
$\tau_2$ must be fixed to $\tau_2={\sqrt3/2}$, the components
of the moduli space $\mathcal{C}^2$ obtained by $D_3(R_1)$ orbifolding
are
$$
\mathcal{C}_{D_3(R_1)-{\rm orb}}^{(\rho_1)}
\cong \R^+\,, \qquad \rho_1\in\left\{0,\half\right\}.
$$
The maximal abelian subgroups of $D_3(R_1)$ are $\Z_3$, and three
order two groups $\{1, R_1\}$,$\{1, A(2\pi/3) R_1 \}$, $\{1, A(4\pi/3)
R_1\}$.  These groups give identical contributions to the partition
function since they are conjugate within $D_3(R_1)$.
Using~(\ref{nabe}) we therefore find
\bea
\label{lattice13}
Z_{D_3(R_1)-{\rm orb}} \left(1/2,\sqrt{3}/2, \rho_1,\rho_2\right) & = &
\frac{1}{6}\left( 3Z_{\Z_3-{\rm orb}} + 3(2Z_{R_1-{\rm orb}} - Z
\right))
\nonumber\\& = &
\frac{1}{2}\left( Z_{\Z_3-{\rm orb}} + 2Z_{R_1-{\rm orb}} - Z \right) ,
\eea
where $Z_{\Z_3-{\rm orb}}$ is  given in~(\ref{lattice3}),
and $Z_{R_1-{\rm orb}}$ is given in~(\ref{lattice7}) or in~(\ref{lattice77}) for 
$\rho_1 = 0$ or $\rho_1 = 1/2$, respectively. 
\subsection{Lattice 14:
                  The $D_3(R_2)$ orbifold}
Lattice 14  $(\tau = e^{2\pi i/3})$ has  a
$D_3(R_2) =\Z_3\cup \{ R_2, A(2\pi/3)R_2, A(4\pi/3)R_2\}$ symmetry.
Analogously to lattice 13 we find
\be\label{lattice14}
Z_{D_3(R_2)-{\rm orb}}\left(\frac 12, \frac{\sqrt{3}}2, \rho_1,\rho_2\right) = 
          \frac{1}{2}\left( Z_{\Z_3-{\rm orb}} + 2Z_{R_2-{\rm orb}} - Z \right) ,
\ee
where $Z_{\Z_3-{\rm orb}}$ is given in~(\ref{lattice3}), and
$Z_{R_2-{\rm orb}}$ is given in ~(\ref{lattice7})
and~(\ref{lattice77}) for $\rho_1 = 0$ or $\rho_1 = 1/2$,
respectively.

From our discussion of lattices 6 and 7 we know that $Z_{R_2-{\rm
orb}}$ is obtained from $Z_{R_1-{\rm orb}}$ by application of
T-duality~(\ref{exchange}) on $\tau$. Using mirror symmetry we see
that we can equally apply T-duality to $\rho$ and find
\begin{eqnarray*}
Z_{D_3(R_2)-{\rm orb}}\left(\frac 12, \frac{\sqrt{3}}2, 0,\rho_2\right)
&=& Z_{D_3(R_1)-{\rm orb}}\left(\frac12, \frac{\sqrt{3}}2, 0,\frac{1}{\rho_2}\right),
\\
Z_{D_3(R_2)-{\rm orb}}\left(\frac12, \frac{\sqrt{3}}2, \frac12,\rho_2\right)
&=& Z_{D_3(R_1)-{\rm orb}}\left(\frac12, \frac{\sqrt{3}}2, \frac12,\frac1{4\rho_2}\right).
\end{eqnarray*}
The above actually is the equation for T-duality on
$\mathcal{C}_{D_3(R_1)-{\rm orb}}^{(\rho_1)}$. In particular, the
$D_3(R_2)$ orbifold procedure does not yield a new component of the
moduli space $\mathcal{C}^2$ but only reproduces
$\mathcal{C}_{D_3(R_1)-{\rm orb}}^{(\rho_1)},
\rho_1\in\{0,{1/2}\}$.

\subsection{Lattice 15: the $D_4$ orbifold}

Lattice 15 $(\tau = i)$ has a $D_4 = \Z_4\cup \{ R_1, A(\pi/2) R_1,
R_2, A(\pi/2) R_2\}$ symmetry.  The maximal abelian subgroups of $D_4$
are $\Z_4$, $D_2 = \{1, A(\pi), R_1, R_2\}$, and $D^\prime_2 = \{1,
A(\pi), A(\pi/2) R_1, A(\pi/2) R_2\}$.  The two order four groups
$D_2$ and $D^\prime_2$ give different contributions to the partition
function, since these groups are not conjugate in $D_4$. The
fundamental cells of lattice 15 we have to pick in order to interprete
them as reflections along the edges of the cell have different
shape. For $D_2$ it is a unit square giving a contribution
$Z_{D_2-{\rm orb}}(\tau=i,\rho)$, whereas for $D_2^\prime$ it is a
rhombus giving a contribution $Z_{D_2-{\rm
orb}}(\tau={1/2}+{i/2},\rho)$.  Note that by~(\ref{lattice8})
for $\rho_1=0$ we have an independent choice of sign for the discrete
torsion parts of $D_2, D_2^\prime$, and for $\rho_1=1/2$ discrete
torsion enters for $D_2$ only.  Using~(\ref{nabe}) and
$\delta,\epsilon\in\{\pm\}$ for the partition function we therefore
get
\bea
\label{lattice15}
Z_{D_4^{\delta \epsilon}-{\rm orb}}(0, 1 , 0,\rho_2) &=&
\frac{1}{2}\Biggl( Z_{\Z_4-{\rm orb}}(0, 1 , 0,\rho_2) +
Z_{D_2^\delta-{\rm orb}}(0,1,0,\rho_2) +
\nonumber\\& & 
	\hphantom{\frac{1}{2}\Biggl(}\!
+ Z_{D_2^\epsilon-{\rm orb}}\left(\frac12, \frac12 , 0,\rho_2\right) -
Z_{\Z_2-{\rm orb}}(0,1,0,\rho_2)\Biggr)\,,
\nonumber\\
Z_{D_4^{\pm}-{\rm orb}}\left(0, 1 , \frac12,\rho_2\right) &=& \frac{1}{2}\Biggl(
Z_{\Z_4-{\rm orb}}\left(0, 1 , \frac12,\rho_2\right) + Z_{D_2^\pm-{\rm
orb}}\left(0,1,\frac12,\rho_2\right) +
\nonumber \\& & 
	\hphantom{\frac{1}{2}\Biggl(}\!
+ Z_{D_2-{\rm orb}}\left(\frac12, \frac12 , 
\frac12,\rho_2\right) - Z_{\Z_2-{\rm orb}}
\left(0,1,\frac12,\rho_2\right)\Biggr),
\qquad~
\eea
where $Z_{\Z_2-{\rm orb}}$, $Z_{\Z_4-{\rm orb}}$ and $Z_{D_2^\pm-{\rm
orb}}$ are given in (\ref{lattice2}),~(\ref{lattice4}),
and~(\ref{lattice8}).  We remark that in case $\rho_1=1/2$ the
$Z_{\Z_4}, Z_{D_2}, Z_{\Z_2}$ parts of~(\ref{lattice15}) always
contribute even numbers of fields with dimensions $h=\bar
h=1/16$. This shows that for the $D_2$ orbifold with $\tau_1=0,
\rho_1=1/2$ we must indeed have $k=2$ in~(\ref{lattice8}).

Since for the $D_2$ orbifold by our discussion of lattices 8 and 9
separate T-duality may be performed on $\tau,\rho$ without changing
the theory, the $D_4$ orbifold conformal field theories form six
families
$$
\mathcal{C}_{D_4^{\delta\epsilon}-{\rm orb}}^{(0)}
\cong\R^+/\Theta\,, \quad \delta,\epsilon\in\{ \pm \}\,,\qquad
\mathcal{C}_{D_4^{\pm}-{\rm orb}}^{(1/2)}
\cong\R^+/\Theta\,.
$$

\subsection{Lattice 16: the $D_4(\mathbb{T}^\prime_R)$ orbifold}

Lattice 16 $(\tau = i)$ has a $D_4(\mathbb{T}^\prime_R) = \Z_4\cup\{
\mathbb{T}^\prime_{R_1}, A(\pi/2)\mathbb{T}^\prime_{R_1},
\mathbb{T}^\prime_{R_2}, A(\pi/2)\mathbb{T}^\prime_{R_2}\}$ symmetry
as defined in~(\ref{symmetries}), and we may set $\rho_1=0$.  The
maximal abelian subgroups of $D_4(T_R^\prime)$ are $\Z_4$,
$D_2(\mathbb{T}^\prime_R)= \{1, A(\pi), \mathbb{T}^\prime_{R_1},
\mathbb{T}^\prime_{R_2} \}$, and $D_2 = \{1, A(\pi), A(\pi/2)
\mathbb{T}^\prime_{R_1}, A(\pi/2)\mathbb{T}^\prime_{R_2}\}$.
Anologously to lattice 15 we find
\bea
\label{lattice16}
Z_{D_4(\mathbb{T}^\prime_R)^\pm-{\rm orb}}(0, 1, 0,\rho_2) &=&
\frac{1}{2}\Biggl( Z_{\Z_4-{\rm orb}}(0, 1, 0,\rho_2) +
Z_{D_2(\mathbb{T}^\prime_R)-{\rm orb}}(0, 1, 0,\rho_2) +
\\& &
	\hphantom{\frac12\Biggl(}\!
 + Z_{D_2^\pm-{\rm orb}}\left(\frac12, \frac12, 0,\rho_2\right) 
- Z_{\Z_2-{\rm orb}}(0, 1, 0,\rho_2)\Biggr) ,\qquad
\nonumber
\eea
where $Z_{\Z_2-{\rm orb}}$, $Z_{\Z_4-{\rm orb}}$,
$Z_{D_2(\mathbb{T}^\prime_R)-{\rm orb}}$ and $Z_{D_2^\pm-{\rm orb}}$
are given in~(\ref{lattice2}),~(\ref{lattice4}),~(\ref{lattice12}),
and~(\ref{lattice8}).  The discussion of the $D_2(T_R^\prime)$
orbifold (lattice 12) shows that T-duality does not map equivalent
$D_4(T_R^\prime)$ orbifold theories onto each other, thus
$$
\mathcal{C}_{D_4(T_R^\prime)^\pm-{\rm orb}}
\cong\R^+\,.
$$

\subsection{Lattice 17: the $D_6$ orbifold}

Lattice 17 $(\tau = e^{2\pi i/3})$ has a $D_6 = \Z_6\cup\{ R_1,
A(\pi/3)R_1, A(2\pi/3)R_1, R_2, A(\pi/3)R_2,$ $A(2\pi/3)R_2\}$
symmetry.  The maximal abelian subgroups of $D_6$ are $\Z_6$, and
three groups of type $D_2$, namely $\{1, A(\pi), R_1,$ $R_2\}$, $\{1,
A(\pi), A(\pi/3)R_1, A(\pi/3)R_2\}$, $\{1,$\linebreak $ A(\pi)$,
$A(2\pi/3)R_1$, $A(2\pi/3)R_2 \}$.  These order four groups give
identical contributions to the partition function since they are
conjugate in $D_6$. This also means that in order for the action of
$D_6$ on the twisted sector to be well defined, discrete torsion must
be the same for all the three of them.  Now by~(\ref{nabe}) the
complete partition function is
\bea
\label{lattice17}
Z_{D_6^{(\pm)}-{\rm orb}} \left(\frac12, \frac{\sqrt{3}}2, \rho_1,
\rho_2\right) &=& \frac{1}{12}\left(6Z_{\Z_6-{\rm orb}} + 
3\left(4Z_{D_2^{(\pm)}-{\rm orb}} - 2Z_{\Z_2-{\rm orb}}\right)\right)
\nonumber\\
&=&\frac{1}{2}\left(Z_{\Z_6-{\rm orb}} + 2Z_{D_2^{(\pm)}-{\rm orb}} -
Z_{\Z_2-{\rm orb}}\right) ,
\eea
where $Z_{\Z_2-{\rm orb}}$, $Z_{\Z_6-{\rm orb}}$ and
$Z_{D_2^{(\pm)}-{\rm orb}}$ are given in
(\ref{lattice2}),~(\ref{lattice5}), and~(\ref{lattice8}).  Analogously
to lattice 15, the three families of $D_6$ orbifold conformal field
theories are
$$
\mathcal{C}_{D_6^\pm-{\rm orb}}^{(0)}
\cong\R^+/\Theta\,, \qquad 
\mathcal{C}_{D_6-{\rm orb}}^{(1/2)}
\cong\R^+/\Theta\,.
$$

\section{Multicritical lines and points}\label{meet}

We now determine all intersections of the 28 nonexceptional components
$\mathcal{C}^{(\bullet)}_{G^{(\bullet)}-{\rm orb}}$ of the moduli
space that we constructed in section~\ref{part}.  We find that all but
four of them can be connected directly or indirectly to the moduli
space $\mathcal{T}^2$ of toroidal theories, and $\mathcal{C}^2$
exhibits a complicated structure with various loops.

The procedure closely follows the proof for the isomorphy of the $c=1$
circle theory at radius $r=2$ to the orbifold theory at radius $r=1$
(see, e.g.,~\cite{har,gin}).  The main idea is to exploit the enhanced
$\SU(2)$ symmetry of the circle theory at radius $r=1$. Namely,
$\SU(2)$ relates two generically different $\Z_2$ actions in this
theory by conjugation. Thus the resulting orbifold theories are
isomorphic. One of them is the circle theory at doubled radius $r=2$,
the other is the ordinary $\Z_2$ orbifold theory at radius $r=1$.

Using results of B.~Rostand's we can show that the generalization of
the above procedure to $c=2$ will suffice to find all intersections of
our $29$ nonexceptional nonisolated components of $\mathcal{C}^2$.
Namely, in~\cite{ro90,ro91} it is shown that every multicritical point
on the moduli space $\mathcal{T}^2$ of toroidal theories is an
orbifold of another toroidal theory with enhanced symmetry. By our
discussion in section~\ref{tori}, we may restrict ourselves to the
study of left-right symmetric orbifolds. In particular, to find all
intersections of $\mathcal{T}^2$ with one of the $28$ nonexceptional
orbifold components it suffices to determine all toroidal theories
with enhanced left and right symmetry (which in the following are
simply called theories with enhanced symmetry) and mod out all
symmetries which are conjugate to some shift on the charge lattice. As
anticipated in~\cite{dijk} each of the toroidal multicritical points
generates a series of further multicritical points or lines, since we
can mod out further symmetries. But even better, this procedure will
lead to the determination of all intersection points: by the
discussion in sections \ref{tori} and~\ref{symm}, all the $28$
nonexceptional components of $\mathcal{C}^2$ are obtained by modding
out solvable groups from toroidal theories. This means that we can
always regain the original toroidal theory by performing another
orbifold procedure. In particular, any intersection point between
nonexceptional nonisolated components of $\mathcal{C}^2$ corresponds
to a multicritical point on $\mathcal{T}^2$.

One can simplify things by stepwise modding out~\cite{dix}: if a
symmetry group $G$ contains a normal subgroup $H$, then the $G$
orbifold conformal field theory $A/G$ of a theory $A$ is isomorphic to
the $G/H$ orbifold conformal field theory of $A/H$.  Moreover, the
$G/H$ action on $A/H$ translates to an action on any other theory
$A^\prime$ which was identified with $A/H$. For $H^\prime\subset G/H,
G^\prime\cong H\times H^\prime$ this leads to possibly new
identifications $A/G^\prime\cong A^\prime/H^\prime$ which need not
correspond to conjugate actions on the original $A$. In $A/H\cong
A^\prime$ we may have gotten rid of all states which the $G^\prime$
action has no consistent conjugate on.

In section~\ref{enhanced} we start by determining all points of
enhanced symmetry in $\mathcal{T}^2$. The idea of proof again is
closely related to the techniques used in~\cite{ro90,ro91}.  In
section~\ref{multtor} we discuss all the multicritical points and
lines obtainable by modding out conjugate $\Z_2$ symmetries of tori
with enhanced $\SU(2)$ symmetry. In sections~\ref{l1l5}--\ref{t2} we
determine all multicritical points and lines obtainable from those
identifications we found in~\ref{multtor} by modding out further
symmetries.  Afterwards (section~\ref{z3mult}) we follow the same
procedure for the $\SU(3)$ torus theory at $\tau=\rho=e^{2\pi i/3}$.
The slightly technical discussion results in a list of all
multicritical points and lines in nonexceptional nonisolated
components of $\mathcal{C}^2$.

We remark that all the identifications below have been confirmed by us
on the level of partition functions numerically.  We will denote the
$G^{(\bullet)}$ orbifold theory of the toroidal theory
$A_T(\tau_1,\tau_2,\rho_1,\rho_2)$ with parameters $(\tau,\rho)$ by
$A_{G^{(\bullet)}-{\rm orb}}(\tau_1,\tau_2,\rho_1,\rho_2)$ in the
following.

\subsection{Points of enhanced symmetry in $\mathcal{T}^2$}\label{enhanced}

Assume that a toroidal conformal field theory with charge lattice 
$\Gamma$ has enhanced symmetry. By 
$\{(\pm p_i,0), (0,\pm p_i^\prime), 
i\in\{ 1,\dots,d\}\}\subset\Gamma$ we denote
the charge vectors corresponding to the additional vertex operators 
of dimensions $(1,0)$ and $(0,1)$, respectively. In particular, 
$|p_i|^2=|p_i^\prime|^2=2$, and since the corresponding vertex operators
are pairwise local, for $i\neq j$ we may assume 
$p_i\cdot p_j = p_i^\prime\cdot p_j^\prime\in\{0,1\}$. Then the
$\R$--span of $\{(p_i,p_i^\prime), i\in\{ 1,\dots,d\}\}\subset\Gamma$
is totally isotropic with respect to the scalar product~(\ref{skp}).
This means that we may choose a geometric interpretation $(\Lambda,B)$
of our toroidal theory such that $p_i=p_i^\prime$ for all $i\in\{ 1,\dots,d\}$
(see~\cite{as96,nawe}). Moreover, by the above restrictions on the scalar
products between the $p_i$, these vectors generate the root lattice
of a simply laced Lie group. Since the rank of this group
can be at most two, the only possible groups are 
$A_2=\SU(3),A_1^2=\SU(2)^2$ or $A_1=\SU(2)$.
If we now write the charge vectors $(p_i,0)$ and $(0,-p_i)$ in the form
(\ref{charges}), we find
$$
\fa i\in\{1,\dots,d\}:\quad \pm p_i = {1\over\sqrt2}\left( \mu_i^\pm -
B\lambda_i \pm \lambda_i \right),\qquad \lambda_i\in\Lambda,
\mu_i^\pm\in\Lambda^\ast\,.
$$
In particular, $2\lambda_i, 2B\lambda_i\in\Lambda^\ast$ for all
$i\in\{1,\dots,d\}$. These conditions suffice to determine all
theories in $\mathcal{T}^2$ with (left-right symmetrically) enhanced
symmetry.

There are two theories with maximally (i.e.\ rank two) enhanced
symmetry, namely the $\SU(2)^2$ torus theory at $\tau=\rho=i$ and the
$\SU(3)$ torus theory at $\tau=\rho=e^{2\pi i/3}$.  Tori with
$\tau=\rho\not\in\{i,e^{2\pi i/3}\}$ and $\tau_1\in\{0,{1/2}\}$
exhibit an enhanced $\SU(2)$ symmetry.

\subsection{Multicritical lines on the torus moduli space $\mathcal{T}^2$:
conjugate $\Z_2$ actions}\label{multtor}

To compare all $\Z_2$ symmetries of the $\SU(2)^2$ torus theory at
$\tau=\rho=i$ we discuss their action on the $(1,0)$ fields. As in
section \ref{tori}, the conserved currents of the generic toroidal
theory are called $j_\mu$. The additional vertex operators of
dimensions $(1,0)$ are denoted $j_\mu^\pm, \mu\in\{1,2\}$, such that
each triple $j_\mu,j_\mu^\pm$ generates an $\SU(2)_1$ Kac--Moody
algebra. Each of these $\SU(2)_1$ Kac--Moody algebras belongs to one
of the $c=1$ factors of the torus theory.  Let us list all $\Z_2$
symmetries with two positive and four negative eigenvalues on the set
of $(1,0)$ fields. By $\widetilde{\Z_2(R)}$ we denote the $\Z_2(R)$
symmetry applied to the torus theory with fundamental cell such that
$\tau=\rho=1/2+i/2$ (remember the phases on Hilbert space ground
states that were discussed for lattice~9):
$$
\begin{array}{lll}
\Z_2 \mbox{ rotational group}:
& j_\mu\mapsto-j_\mu,& j_\mu^\pm\leftrightarrow j_\mu^\mp\,,
\\[4pt]
\mbox{shift orbifold by } \delta^\prime={1\over2}\left({1\atop1}\right):\;
& j_\mu\mapsto j_\mu,& j_\mu^\pm\mapsto -j_\mu^\pm\,, 
\\[4pt]
\widetilde{\Z_2(R_1)}:
&j_1\mapsto j_1, j_2\mapsto -j_2,\;\;
&j_1^\pm\mapsto -j_1^\pm, j_2^+\leftrightarrow j_2^-\,,
\\[4pt]
\Z_2(T_{R_1}):
&j_1\mapsto j_1, j_2\mapsto -j_2,
&j_1^\pm\mapsto -j_1^\pm, j_2^+\leftrightarrow j_2^-\,.
\end{array}
$$
None of the above symmetries mixes currents from different $c=1$
factors of the torus theory or $j_\mu$ with $j_\mu^\pm$
currents. Moreover, their eigenvalue spectrum is identical on each
$c=1$ factor, so we may use the corresponding $c=1$ result to show
that the four $\Z_2$ orbifolds by the above listed symmetries give
isomorphic theories when applied to the $\SU(2)^2$ theory. This
generates a quadrucritical point.  The shift orbifold by the half
lattice vector $\delta^\prime$, as usual, results in a torus theory
with additional generator $\delta^\prime$ of the lattice and half
volume and B-field ($A_T(0, 1, 0, 2)=A_T(0, 1, 0, 1/2)$ by
T-duality):
\begin{quadru}
A_T(0, 1, 0, 2)&=&A_{T_{R}-{\rm orb}}(0, 1, 0, 1)
\nonumber\\
&=&A_{\Z_2-{\rm orb}}(0, 1, 0, 1)=A_{R-{\rm orb}}
\left(\frac12,\frac12,\frac12,\frac12\right).
\label{t_z2}
\end{quadru}
The equality $A_T(0, 1, 0, 2)=A_{\Z_2-{\rm orb}}(0, 1, 0, 1)$ has
already been proven in~\cite{ken}, both on the level of partition
function and operator algebra.

The above quadrucritical point turns out to actually be the
intersection of four bicritical lines.  First consider the family of
torus theories at parameters $\tau=\rho=it, t\in\R^+$ which decompose
into tensor products of two $c=1$ circle theories at radii $r=1$ and
$r^\prime=t$, respectively.  For all values of $t$ the first factor
possesses an $\SU(2)$ symmetry.  Since the actions of $T_{R_2}$ and
the shift by $\delta^\prime={1\over2}\left({1\atop t}\right)$ only
differ on this first factor, where they are generally conjugate by the
$\SU(2)$ symmetry, we find ($A_T(1/2, t/2,0, t/2)=A_T(0, t/2, 1/2,
t/2)$ by mirror symmetry)
\bl{t_z2tr}
\fa t\in\R^+:\quad 
A_T\left(0, \frac t2, \frac12, \frac t2\right)= 
A_{T_{R_2}-{\rm orb}}(0, t, 0, t) \,,
\el
and analogously 
\bl{z2tr_z2}
\fa t\in\R^+:\quad A_{T_{R_1}-{\rm orb}}(0, t, 0, t) 
= A_{\Z_2-{\rm orb}}(0, t, 0, t)\,.
\el
Next consider the family of toroidal theories at parameters
$\tau=\rho=1/2+it$, $t\in\R^+$. We also have a generic $\SU(2)\times
\U(1)$ symmetry for this family. Inspection of the charge lattice
shows that as before we have conjugate $\Z_2$ symmetries now giving
bicritical lines
\bls
\fa t\in\R^+:\quad
A_{\Z_2-{\rm orb}}\left(\frac12, t, \frac12, t\right) 
&=& A_{R_1-{\rm orb}}\left(\frac12, t, \frac12, t\right),
\llabel{z2rhh_z2}\\
A_{R_2-{\rm orb}}\left(\frac12,t, \frac12, t\right) 
&=& A_T\left(0, 2t, \frac14, \frac t2\right) .
\llabel{t_z2rhh}
\els
There are two more $\Z_2$ symmetries which are conjugate on the entire
family of toroidal theories with parameters $\tau=\rho=it, t\in\R^+$
by $\SU(2)$ symmetry on the first factor. They have four positive and
two negative eigenvalues on $(1,0)$ fields:
$$
\begin{array}{llll}
\Z_2(R_2):&j_1\mapsto -j_1\,,\quad j_2\mapsto j_2\,,\quad
&j_1^+\leftrightarrow j_1^-\,,\quad &j_2^\pm\mapsto j_2^\pm\,,
\\[4pt]
\mbox{shift orbifold by } \delta_1={1\over2}\left({1\atop0}\right):\;
& j_\mu\mapsto j_\mu\,,\quad& j_1^\pm\mapsto -j_1^\pm\,,\quad &j_2^\pm\mapsto j_2^\pm\,.
\end{array}
$$
In particular,
\bl{t_z2r00}
\fa t\in\R^+:\quad
A_{R_2-{\rm orb}}(0,t, 0,t) =  A_T\left(0,\frac t2, 0,2t\right).
\el
We remark that $\Z_2(R)$ applied to the theory with fundamental cell such
that $\tau=1/2+i/2, \rho=i$ has three positive and three negative eigenvalues
on the set of $(1,0)$ fields. Hence it is not conjugate to any other
crystallographic symmetry of $A_T(0,1,0,1)$.

\subsection{Series of multicritical lines and points obtainable from 
(L1) and (L5)}\label{l1l5}

We are now going to mod out further symmetries on both sides of the
equalities obtained above. The main problem is to find the correct
translation for the action of a symmetry from one model to the
other. The simplest case is~(\ref{t_z2r00}) from which we mod out
$R_1$ on both sides. Because all the symmetries used so far respect
the factorization of $A_T(0,t,0,t)$ into a tensor product of two
circle theories and commute, we directly get
\bl{z2r00_d200+}
\fa t\in\R^+:\quad A_{D_2^+-{\rm orb}}(0, t, 0, t)  =  
A_{R_1-{\rm orb}}\left(0, \frac t2, 0, 2t\right).
\el
Note that by mirror symmetry and T--duality~(\ref{exchange}) we have
$A_{R_1-{\rm orb}}(0, 2, 0, 1/2)=A_{R_2-{\rm orb}}(0, 2, 0, 2)$, hence
the above multicritical line and the one found in (\ref{t_z2r00})
intersect in a tricritical point:
\bt{z2r00_d200+_t} A_{D_2^+-{\rm orb}}(0, 1, 0, 1)  =  
A_{R_1-{\rm orb}}\left(0, \frac 12, 0, 2\right) = A_T(0, 1, 0, 4)\,.
\et
We now systematically mod out all symmetries of the torus theory
$A_T(0, t/2, 0, 2t)$ in~(\ref{t_z2r00}). The procedure is similar in
all cases, namely, the charge lattices of the underlying toroidal
theories on both sides of an identification must be determined, as
well as twisted ground states, if present. After having performed a
state by state identification, symmetries can be translated from one
side to the other. This way the details which we partly omit in the
proofs below can easily be filled.

As to~(\ref{t_z2r00}), by~(\ref{group}) the actions we can generically
mod out on the torus theory $A_T(0, t/2, 0, 2t)$ are $\Z_2, \Z_2(R),
\Z_2(T_R), D_2^\pm, D_2(T_R)$ and $D_2(T_R^\prime)$.  At $t=2$ one has
additional $\widetilde{Z_2(R)}$ and $\Z_4$ actions which give no new
identifications, though.

Modding out by $\Z_2(R_1)$ gives the bicritical
line~(\ref{z2r00_d200+}) as discussed above. The reflection $R_2$ on
the torus side acts as a shift by
$\delta_1={1\over2}\left(1\atop0\right)$ on the underlying torus
theory of $A_{R_2}(0,t,0,t)$ leading to a trivial identity. The
symmetry $T_{R_1}$ applied to the torus side differs in its action
from $R_1$ by additional signs on those vertex operators (of lowest
dimension) in $A_T(0, t/2, 0, 2t)$ which correspond to twisted ground
states in $A_{R_2}(0,t,0,t)$. Therefore, comparison
with~(\ref{z2r00_d200+}) shows
\bl{z2tr00_d200-}
\fa t\in\R^+:\quad A_{D_2^--{\rm orb}}(0, t, 0, t)  =  
A_{T_{R_1}-{\rm orb}}\left(0, \frac t2, 0, 2t\right).
\el
Modding out by $T_{R_2}$ instead of $T_{R_1}$ again gives a trivial
identity, since $T_{R_2}$ acts on the underlying torus of $A_{R_2-{\rm
orb}}(0,t,0,t)$ by the shift $T_{\delta_1}$. Note that a comparison
of~(\ref{z2tr00_d200-}) with~(\ref{z2r00_d200+}) gives a fairly
natural explanation for the additional degree of freedom we have due
to discrete torsion.  Since $A_{T_{R_1}-{\rm
orb}}(0,1/2,0,2)=A_{T_{R_2}-{\rm orb}}(0,2,0,2)$ by T--duality (see
the discussion of lattice 10), the multicritical lines
(\ref{z2tr00_d200-}) and (\ref{t_z2tr}) intersect in a tricritical
point:
\bt{t_d200-}
A_{D_2^--{\rm orb}}(0, 1, 0, 1)=A_{T_{R_2}-{\rm orb}}
(0, 2, 0, 2)=A_T\left(0, 1, \frac12, 1\right).
\et
Next, we mod out the ordinary $\Z_2$ action on~(\ref{t_z2r00}).  The
multicritical line~(\ref{t_z2r00}) can also be written as
$A_{\widehat{T}_{R_2}-{\rm orb}}(0,t, 0,t) = A_T(0,t/2, 0,2t)$.
Recall from~(\ref{circle2}) that $A_T(0,t, 0,t)$ as well as $A_T(0,2t,
0,t/2)$ are tensor products of circle theories at radii
$r=1$, $r^\prime=t$ and $r=2$, $r^\prime=t$, respectively. Now consider the
residual action of $D_2(T_R)$ of the original torus theory $A_T(0,t,
0,t)$ on the orbifoldized theory $A_{\widehat{T}_{R_2}-{\rm orb}}(0,t,
0,t)$ and note that it acts as ordinary $\Z_2$ on the invariant
sector.  The twisted ground states of the first circle factor are
interchanged, so all in all we get an ordinary $\Z_2$ action on the
torus theory $A_T(0,t/2, 0,2t)$. This yields
\bl{d2tr_z2}
\fa t\in\R^+:\quad A_{D_2(T_R)-{\rm orb}}( 0, t, 0, t) = 
A_{\Z_2-{\rm orb}}\left(0, \frac t2, 0, 2t\right).
\el
By analogous arguments one finds that modding out~(\ref{t_z2tr}) by
$\Z_2$ on the torus side yields
\bl{d2trp_z2}
\fa t\in\R^+:\quad A_{\Z_2-{\rm orb}}\left(0, \frac t2, \frac 12,
\frac t2\right)=A_{D_2(\mathbb{T}^\prime_R)-{\rm orb}}( 0, t, 0, t)\,.
\el
As mentioned above, $R_2$ applied to the torus theory
$A_T(0,t/2,1/2,t/2)$ acts as shift $T_{\delta_1}$ on the underlying
torus theory of $A_{{R_2}-{\rm orb}}(0,t, 0,t)$.  Applying this to the
bicritical line~(\ref{d2tr_z2}), if $R_2$ acts with positive sign on
the $\Z_2$ twisted ground states of the right hand side we obtain a
trivial identity. On the other hand, if we use negative discrete
torsion on the right hand side we find
\bl{d200-_d2tr}
\fa t\in\R^+:\quad A_{D_2(T_R)-{\rm orb}}\left(0, \frac t2, 0,
2t\right) = A_{D_2^--{\rm orb}}\left( 0, \frac t2, 0, 2t\right).
\el
Note that the bicritical lines~(\ref{z2tr00_d200-})
and~(\ref{d200-_d2tr}) intersect in a tricritical point which can be
interpreted as the result of modding out~(\ref{z2r00_d200+_t}) by
$T_{R_1}$:
\be
\label{z2tr00_d200-_d2tr}
A_{T_{R_1}-{\rm orb}}(0,1,0,4) =
A_{D_2^--{\rm orb}}( 0, 2, 0, 2) = A_{D_2(T_R)-{\rm orb}}\left(0,
\frac 12, 0, 2\right).
\ee 
To mod out (\ref{t_z2r00}) by $D_2(T_R^\prime)$ on the torus side
amounts to modding out~(\ref{z2tr00_d200-}) by $T_{R_2}^\prime$ which
acts as shift $T_{\delta^\prime}, \delta^\prime = {1\over2}\left(
{1\atop t} \right)$ on the underlying torus of $A_{D_2^--{\rm
orb}}(0,t,0,t)$. Thus
\bl{d2h0-_d2trp}
\fa t\in\R^+:\quad
A_{D_2^--{\rm orb}}\left( \frac 12, \frac t2, 0, \frac t2\right)
= A_{D_2(T_R^\prime)-{\rm orb}}\left(0, \frac t2, 0, 2t\right),
\el
Note that because of T--duality $A_{D_2(T_R^\prime)-{\rm
orb}}(0,2,0,2)=A_{D_2(T_R^\prime)-{\rm orb}}(0,1/2,0,2)$ as discussed
for lattice 12, so~(\ref{d2h0-_d2trp}) intersects~(\ref{d2trp_z2}) in
a tricritical point which can be understood as the result of modding
out~(\ref{t_d200-}) by $\Z_2$:
\bt{d200-_d2trp_z2}
A_{D_2^--{\rm orb}}\left( \frac 12, \frac 12, 0, \frac 12\right) 
= A_{D_2(T_R^\prime)-{\rm orb}}\left(0, \frac 12, 0, 2\right)
= A_{\Z_2}\left(0,1,\frac 12,1\right).
\et
We now turn to a systematic discussion of intersection lines and
points obtained from (\ref{t_z2tr}). From $A_T(0,t/2,1/2,t/2)$ we can
generically mod out $\Z_2,\Z_2(R)$ and $D_2^\pm$. The additional
symmetries for $t=1$ and $t=2$ produce nothing new.

Modding out by the ordinary $\Z_2$ action on the torus side gives
the bicritical line~(\ref{d2trp_z2}), as was mentioned above.
We claim that the result of modding out a $\Z_2(R_1)$ action leads to the
bicritical line 
\bl{z2rh0_d2tr}
\fa t\in\R^+:\quad A_{R_1-{\rm orb}}\left( 0, \frac t2, \frac 12,
\frac t2\right) = A_{D_2(T_R)-{\rm orb}}\left(0, \frac 1t, 0, t\right).
\el
Actually, the slightly surprising parameters on the right hand side
are due to an apparent asymmetry in the definition of
$D_2(T_R)=\{1,A(\pi),T_{R_1},\widehat{T}_{R_2}\}$. If we use
$\widehat{D_2(T_R)}=\{1,A(\pi),T_{R_2},\widehat{T}_{R_1}\}$ instead,
then by T--duality (see the discussion of lattice 11) the parameters
on the right hand side of~(\ref{z2rh0_d2tr}) are $(0,t,0,t)$. Our
claim thus amounts to the fact that $R_1$ as applied to $A_T(0, t/2,
1/2, t/2)$ induces an ordinary $\Z_2$ action (or equivalently
$\widehat{T}_{R_1}$) on $A_{T_{R_2}-{\rm orb}}(0,t,0,t)$. For the
$(1,0)$ fields this is easy to check: $R_1$ leaves one of the abelian
currents of the torus theory invariant and multiplies the other by
$-1$. So do $\Z_2$ and $\widehat{T}_{R_1}$ on $A_{T_{R_2}-{\rm
orb}}(0,t,0,t)$, where the $T_{R_2}$ invariant generic abelian current
of the underlying torus theory is multiplied by $-1$, and the
$T_{R_2}$ invariant combination of vertex operators remains invariant.
To give a full proof for~(\ref{z2rh0_d2tr}), note that the charge
lattice of $A_T( 1/2, t/2,0, t/2)$ by~(\ref{char}) is generated by
vectors
\begin{eqnarray*}
&&
(p,\bar{p}) \in\left\{
{ {1\over\sqrt2}}\left( \pmatrix{1\cr -{1/ t}}, 
\pmatrix{1\cr-{1/t}}\right),
{1\over\sqrt2}\left( \pmatrix{1\cr 0}, 
\pmatrix{-1\cr 0}\right),\right.
\\&&
	\hphantom{(p,\bar{p}) \in\Bigl\{}
\left.{1\over\sqrt2}\left( \pmatrix{0\cr {2/ t}}, 
\pmatrix{0\cr{2/t}}\right),
{1\over\sqrt2}\left(\pmatrix{1/2\cr t/2}, 
\pmatrix{-1/2\cr -t/2}\right)\right\}.
\end{eqnarray*}
The four vertex operators of dimension ${1\over4}(1+{1/ t^2})$
given by
$$
e^{{i\epsilon\over\sqrt2}(\varphi^1+\delta\varphi^2/t)}
e^{{i\epsilon\over\sqrt2}(\bar\varphi^1+\delta\bar\varphi^2/t)}\,,
\qquad\epsilon,\delta\in\{\pm1\}
$$
correspond to the following $T_{R_2}$ invariant vertex operators of
$A_T(0,t,0,t)$:
$$
e^{{i\epsilon\over\sqrt2}(\varphi^1_\prime+\varphi^2_\prime/t)}
e^{{i\epsilon\over\sqrt2}(\delta\bar\varphi_\prime^1+\bar\varphi_\prime^2/t)}
-e^{{i\epsilon\over\sqrt2}(-\varphi^1_\prime+\varphi^2_\prime/t)}
e^{{i\epsilon\over\sqrt2}(-\delta\bar\varphi_\prime^1+\bar\varphi_\prime^2/t)}\,,
\qquad\epsilon,\delta\in\{\pm1\}
$$
(see~(\ref{char}) to determine the charge lattice of $A_T(0,t,0,t)$;
$\varphi_\prime^\mu$ denote the bosonic fields in this torus theory to
distinguish them from $\varphi^\mu$ on $A_T( 0, t/2,1/2, t/2)$).  Both
$R_1$ on $A_T(1/2, t/2 , 0, t/2)$ and $\Z_2$ and $\widehat{T}_{R_1}$
on $A_{T_{R_2}-{\rm orb}}(0,t,0,t)$ pairwise interchange these vertex
operators. The four vertex operators of dimension
${1\over16}(1+{t^2})$ given by
$$
e^{{i\over2\sqrt2}(\epsilon\varphi^1+\delta t\varphi^2)}
e^{-{i\over2\sqrt2}(\epsilon\bar\varphi^1+\delta t\bar\varphi^2)}\,,
\qquad\epsilon,\delta\in\{\pm1\}
$$
correspond to the twisted ground states on $A_{T_{R_2}-{\rm
orb}}(0,t,0,t)$, both being pairwise interchanged by $R_1$ on
$A_T(1/2, t/2, 0, t/2)$ and $\Z_2$ and $\widehat{T}_{R_1}$ on
$A_{T_{R_2}-{\rm orb}}(0,t,0,t)$ as well. This proves
(\ref{z2rh0_d2tr}). Modding out $R_2$ instead of $R_1$ gives the same
result, up to T--duality.  Note that the
point~(\ref{z2tr00_d200-_d2tr}) actually lies on~(\ref{z2rh0_d2tr}),
hence we have found another quadrucritical point:
\begin{quadru}
A_{T_{R}-{\rm orb}}(0,1,0,4) 
&=& A_{D_2^--{\rm orb}}( 0, 2, 0, 2)
\nonumber\\ 
&=& A_{D_2(T_R)-{\rm orb}}\left(0, \frac 12, 0, 2\right)
= A_{R-{\rm orb}}\left(  0, 1,\frac 12, 1\right).
\label{z2tr00_d200-_d2tr_z2r}
\end{quadru}
Moreover,~(\ref{z2rh0_d2tr}) intersects the bicritical lines
(\ref{z2tr_z2}) and (\ref{d2tr_z2}), so there is another
quadrucritical point:
\begin{quadru}
A_{R-{\rm orb}}\left(\frac 12, \frac 12, 0, 2\right) &=& 
A_{D_2(T_R)-{\rm orb}}(0, 1, 0, 1)
\nonumber\\
&=&A_{\Z_2-{\rm orb}}(0, 2, 0, 2) = A_{T_{R_1}-{\rm orb}}(0, 2, 0, 2).
\label{z2rh0_z2}
\end{quadru}
We proceed with the above reasoning to see that the $\Z_2$ action on
$A_T( 0, t/2,1/2, t/2)$ translates to a $T_{R_1}^\prime$ action on
$A_{T_{R_2}-{\rm orb}}(0,t,0,t)= A_{T_{R_2}^\prime-{\rm
orb}}(0,t,0,t)$ (this is the proof of~(\ref{d2trp_z2})). Therefore, to
determine the action induced by $D^+_2$ on $A_T( 0, t/2,1/2, t/2)$, we
note that on $A_{T_{R_2}-{\rm orb}}(0,t,0,t)$ the additional symmetry
to mod out compared to~(\ref{z2rh0_d2tr}) on the underlying torus
theory $A_T(0,t,0,t)$ is the combination $T_{R_1}^\prime
\widehat{T}_{R_1}$, i.e.\ a shift by
$\delta_1={1\over2}\left(1\atop0\right)$. Moreover, the $\Z_2$ twisted
ground states in $A_{\Z_2-{\rm orb}}(0,t/2,1/2,t/2)$ are given by
vertex operators which are $\widehat{T}_{R_1}$ invariant, and
therefore
\bl{d200+_d2tr}
\fa t\in\R^+:\quad A_{D_2^+-{\rm orb}}\left(0, \frac t2, \frac 12,
\frac t2\right) = A_{D_2(T_R)-{\rm orb}}\left(0, \frac 2t, 0, 2t\right).
\el
This can also be seen by applying $R_1$ to $A_{\Z_2-{\rm
orb}}(0,t/2,1/2,t/2)$ in (\ref{d2trp_z2}).  Modding out the $D_2^-$
action on the torus side analogously gives (\ref{d2h0-_d2trp}), again.
Note that the bicritical line~(\ref{d200+_d2tr}) intersects
(\ref{d2tr_z2}) and~(\ref{d200-_d2tr}), so we have found two more
tricritical points:
\bt{d2h0+_d2tr_z2}
A_{D_2^+-{\rm orb}}\left(0, \frac 12,\frac 12, \frac 12\right) 
= A_{D_2(T_R)-{\rm orb}}(0, 2, 0, 2)
= A_{\Z_2-{\rm orb}}(0,1,0,4)\,,
\vspace*{-.8em}
\et
\bt{d2h0+_d2tr_d200-}
A_{D_2^+-{\rm orb}}\left(0, 1,\frac 12, 1\right) = 
A_{D_2(T_R)-{\rm orb}}(0, 1, 0, 4)
= A_{D_2^--{\rm orb}}(0, 1, 0, 4)\,.
\et

\subsection{Series of multicritical lines and points obtainable from 
(L2)-(L4)}\label{l24}

To gain further identifications from~(\ref{z2tr_z2}) we can only mod
out further symmetries of the underlying torus theory
$A_T(0,t,0,t)$. If we add generators of order four we only get trivial
identities. An action of $\Z_2(R)$ type basically acts as a shift on
the $A_{T_{R_1}}(0,t,0,t)$ theory, so we arrive at the bicritical
lines (\ref{z2r00_d200+}) and~(\ref{z2tr00_d200-}) again. All other
symmetries give trivial identities.

Next we consider~(\ref{z2rhh_z2}). The symmetries we can generically
mod out are $\Z_2,\Z_2(R)$ and $\Z_2(T_R)$, all giving trivial
identities.  For $t=\sqrt3/2$ we can mod out additional symmetries
containing a $\Z_3$ action, but this does not produce anything new.
For the special value $t=1/2$, where we have $A_{\Z_2-{\rm
orb}}(0,1,0,1) = A_{R_1-{\rm orb}}(1/2, 1/2, 1/2, 1/2)$ all but the
modding out of $T_{R_1}^\prime$ give trivial identities as well. The
symmetry $T_{R_1}^\prime$ multiplies both $\Z_2$ invariant $(1,0)$
fields in $A_{\Z_2-{\rm orb}}(0,1,0,1)$ by $-1$, and the generators of
the invariant part of the $A_T(0,1,0,1)$ charge lattice are pairwise
interchanged. The same is true for the $\Z_2$ twisted ground
states. We claim that this translates to an $R_2$ action on
$A_{R_1-{\rm orb}}(1/2, 1/2, 1/2, 1/2)$. Namely, as a result of the
discussion for lattice 7 we found that on $A_T(1/2,1/2,1/2,1/2)$ the
action of $D_2$ leaves invariant none of the combinations of vertex
operators of dimensions $(1,0)$. The respective $(1/8,1/8)$ and
$(1/2,1/2)$ fields in $A_{R_1-{\rm orb}}(1/2, 1/2, 1/2, 1/2)$ are also
pairwise interchanged, thus
$$
A_{D_2(T_R^\prime)-{\rm orb}}(0, 1, 0, 1)
= A_{D_2-{\rm orb}}\left(\frac 12, \frac 12, \frac 12, \frac 12\right).
$$
By~(\ref{d2trp_z2}) and 
$A_{\Z_2-{\rm orb}}(0, 1/2, 1/2, 1/2)=A_{\Z_2-{\rm orb}}(0, 1, 0, 2)$ 
we see that we have actually found a tricritical point on a bicritical line:
\bt{d2hh_z2_d2trp}
A_{D_2-{\rm orb}}\left(\frac 12, \frac 12, \frac 12, \frac 12\right) = 
A_{\Z_2-{\rm orb}}(0, 1, 0, 2)
=A_{D_2(T_R^\prime)-{\rm orb}}(0, 1, 0, 1).
\et
We remark that the above can be seen more directly by showing that
in the notation of section~\ref{multtor} the groups
$\widetilde{\Z_2(R_1)}\times\widetilde{\Z_2(R_2)}, 
\Z_2\times\Z_2(T_{\delta^\prime})$ and $D_2(T_{R}^\prime)$
are conjugate symmetry groups of type $D_2$ of the $\SU(2)^2$ torus
theory.

In the discussion of lattice 15 we found that $D_4^\pm$ acting
on $A_T(0,1, 1/2, 1/2)$ has a subgroup $D_2^\prime\subset D_4^\pm$
which effectively acts on 
$A_T(1/2,1/2, 1/2, 1/2)=A_T(0,1,0,1)$. By the above this 
is conjugate to the $D_2(T_R^\prime)$ action on
$A_T(0,1,0,1)$, where $D_2(T_R^\prime)\subset D_4^\pm(T_R^\prime)$
generically exactly gives the distinction between 
$D_4^\pm(T_R^\prime)$ and $D_4^\pm$. This means
\begin{equation}
\label{d4h+_d4tr+}
A_{D_4^+-{\rm orb}}\left(0, 1, \frac 12, \frac 12\right) 
= A_{D_4(T_R^\prime)^+-{\rm orb}}(0, 1, 0, 1)\,,
\end{equation}\vspace*{-1em}
\bd{d4h-_d4tr-}
A_{D_4^--{\rm orb}}\left(0, 1, \frac 12, \frac 12\right) = 
A_{D_4(T_R^\prime)^--{\rm orb}}(0, 1, 0, 1)\,.
\ed

Let us now turn to the discussion of~(\ref{t_z2rhh}).
Generically, we can only mod out a $\Z_2$ action on $A_T(0,2t,1/4,t/2)$.
This leads to another bicritical line:
\bl{d2hh_z2}
\fa t\in\R^+:\quad
A_{D_2-{\rm orb}}\left( \frac 12, t, \frac 12, t\right) = 
A_{\Z_2-{\rm orb}}\left(0, 2t, \frac 14, \frac t2\right),
\el
as follows directly from~(\ref{z2rhh_z2}) and~(\ref{t_z2rhh}).
Note that~(\ref{d2hh_z2}) intersects the bicritical line~(\ref{d2trp_z2})  
in~(\ref{d2hh_z2_d2trp}).

We can mod out additional symmetries of~(\ref{t_z2rhh}) at special values
of $t$, namely if $\rho=1/4+it/2$ is equivalent to $\rho^\prime$ with
$\rho^\prime_1\in\{0,1/2\}$ by M\o bius transformations. This is true for
$t\in\{ 1/2, \sqrt3/2, \sqrt7/2, \sqrt{5/12}, \sqrt{3/20}, \sqrt{1/28}\}$,
but only for $t=\sqrt3/2$ we produce a new identification by our methods.
Here,~(\ref{t_z2rhh}) gives 
$A_{R_2-{\rm orb}}(1/2,\sqrt3/2,1/2,\sqrt3/2)=A_T(0,\sqrt3,0,\sqrt3)$, and the
torus theory decomposes into a tensor product of two $c=1$ circle theories
at radii $r=1$ and $r^\prime=\sqrt3$, respectively. The latter only contains
one $(1,0)$ field which is identified with the vertex operator
$e^{i\sqrt{2/3}\,\varphi_1}e^{i\sqrt{2/3}\,\bar\varphi_1}
+ e^{-i\sqrt{2/3}\,\varphi_1}e^{-i\sqrt{2/3}\,\bar\varphi_1}$
in the $A_{R_2-{\rm orb}}(1/2,\sqrt3/2,1/2,\sqrt3/2)$ model. The $\SU(2)$ generators
of the first circle factor are identified with the two other $R_2$ invariant 
vertex operators and the abelian current $j_2$ of 
$A_T(1/2,\sqrt3/2,1/2,\sqrt3/2)$. The only symmetry we can mod out to find 
a new identification is $T_{R_1}$. Then by definition, of the $(1,0)$ fields
on the torus side only one is invariant, namely the abelian current of the
first factor theory. The same is true for the $R_1$ action on 
$A_{R_2-{\rm orb}}(1/2,\sqrt3/2,1/2,\sqrt3/2)$, where only one combination of vertex
operators is invariant. Actually, the actions match entirely, showing
\bd{d2hh_d2tr}
A_{D_2-{\rm orb}}\left(\frac 12,\frac{\sqrt3}2,\frac 12,\frac{\sqrt3}2\right)
= A_{T_{R_1}-{\rm orb}}(0,\sqrt3,0,\sqrt3).
\ed
\subsection{Series of multicritical points obtainable from 
(Q1)}\label{q1}
The identifications in section~\ref{multtor} we have not yet used by
our discussions of the bicritical lines~(\ref{t_z2tr})-(\ref{t_z2r00})
are $A_T(0,1,0,2)=A_{\Z_2-{\rm orb}}(0,1,0,1)$ and $A_{T_{R_1}-{\rm
orb}}(0,1,0,1)=A_{R-{\rm orb}}(1/2,1/2,1/2,1/2)$, taken
\pagebreak[3]
from~(\ref{t_z2}).  In the latter case we can mod out additional
symmetries on the underlying tori, but this produces no new
identifications.  Namely, the ordinary $\Z_2$ action applied to the
left hand side gives the identification $A_{D_2(T_R)-{\rm
orb}}(0,1,0,1)$ $=A_{R_1-{\rm orb}}(0,1/2,1/2,1/2)$
on~(\ref{z2rh0_d2tr}), and $\Z_2$ applied to the right hand side gives
$A_{D_2(T_R^\prime)-{\rm orb}}(0,1,0,1)=A_{D_2-{\rm
orb}}(1/2,1/2,1/2,1/2)$, see (\ref{d2hh_z2_d2trp}).  In fact, by the
discussion at the beginning of the section we know that it suffices to
mod out further symmetries of identities that contain toroidal
theories.

We are now going to mod out further symmetries on both sides of the
equality $A_{\Z_2-{\rm orb}}(0,1,0,1)=A_T(0,1,0,2)$.  
We mostly use the description in terms of the
toroidal theory $A_T(0,1,0,2)$, which by~(\ref{char}) has charge vectors
\be\label{chch}
\mpbar p = {1\over2} \left\{
\left( n_2\atop n_1 \right) \pm 2 \left( m_2\atop m_1 \right) \right\},
\quad m_i,n_i\in\Z.
\ee
On the $A_{\Z_2-{\rm orb}}(0,1,0,1)$ side, the torus currents $J_1,J_2$ of 
$A_T(0,1,0,2)$ are $\Z_2$ invariant combinations of vertex operators
with dimensions $(h,\bar h)=(1,0)$  in the two 
$c=1$ factors of $A_T(0,1,0,1)$. 
The states 
$|0,0,\pm1,0\rangle$, $|0,0,0,\pm1\rangle$ in $A_T(0,1,0,2)$
by~(\ref{chch}) correspond to the $(1/8,1/8)$ fields of the theory
and therefore are identified with the
four twisted ground states of the $\Z_2$ orbifold $A_{\Z_2-{\rm orb}}(0,1,0,1)$.
Further generators of the Hilbert space of $A_T(0,1,0,2)$ are vertex operators
corresponding to
$|\pm1,0, 0,0 \rangle$, $|0,\pm1,0,0\rangle$ which are identified with the
$\Z_2$ invariant combinations of vertex operators with
dimensions $(h,\bar h)=(1/2,1/2)$ of the $A_T(0,1,0,1)$ side.
These do not live in one of the separate factor theories.

The $\Z_2$ action on $A_T(0,1,0,2)$ induces a $\widetilde{\Z_2(R)}$ 
action on the underlying torus of $A_{\Z_2-{\rm orb}}(0,1,0,1)$, and we 
arrive at $A_{D_2-{\rm orb}}(1/2, 1/2, 1/2, 1/2)=A_{\Z_2-{\rm orb}}(0, 1, 0, 2)$
reproducing part of~(\ref{d2hh_z2_d2trp}).
The $R_1$ action on $A_T(0,1,0,2)$  translates to $A_{\Z_2-{\rm orb}}(0,1,0,1)$
in the following way: among the $(1,0)$ fields in $A_{\Z_2-{\rm orb}}(0,1,0,1)$
only the combination in
the first factor of $A_T(0,1,0,1)$ is invariant;
two of the twisted ground states of the $\Z_2$ orbifold are exchanged,
whereas two of them are fixed. 
Among the $(1/2,1/2)$ fields, again two are fixed and two are exchanged; 
this is just the $R_1$ action on $A_{\Z_2-{\rm orb}}(1/2,1/2,0,1)$,
hence 
\bd{d2h0+_z2r00}
A_{D^{+}_2-{\rm orb}}\left(\frac 12, \frac 12, 0, 1\right) = 
A_{R-{\rm orb}}(0, 1, 0, 2)\,.
\ed
If we combine the $\Z_2$ and $\Z_2(R)$ actions on $A_T(0,1,0,2)$, the
$\Z_2$ now will act as a shift on the underlying torus of
$A_{\Z_2-{\rm orb}}(0,1,0,1)$. It is easier to understand the
resulting identification by considering the $\Z_2$ orbifold theory
$A_{\Z_2-{\rm orb}}(0,1,0,2)$.  $T_{R_1}^\prime$ acts on $A_{\Z_2-{\rm
orb}}(0,1,0,2)$ by pairwise interchanging the $\Z_2$ twisted ground
states and multiplying the $\Z_2$ invariant vertex operators of
dimensions $(1/8,1/8)$ in $A_T(0,1,0,2)$ by $-1$. On the other hand,
$R_1$ with negative discrete torsion will multiply the two
$T_{R_1}^\prime$ invariant twisted ground state combinations by $-1$
but leave invariant the two $\Z_2$ invariant $(1/8,1/8)$ fields of
$A_T(0,1,0,2)$.  These $\Z_2$ actions are conjugate, since the action
on the invariant $\Z_2$ twisted ground state combinations of
$A_{\Z_2-{\rm orb}}(0,1,0,1)=A_T(0,1,0,2)$ is merely exchanged with
that on two combinations of twisted ground states of $A_{\Z_2-{\rm
orb}}(0,1,0,2)$.  This again is possible because of the $c=1$
identification between the circle theory at radius $r=1$ and the
orbifold theory at radius $r=2$. In summary,
\bd{d200-_d2trpb}
A_{D_2(T_R^\prime)-{\rm orb}}(0, 1, 0, 2) = A_{D_2^--{\rm orb}}(0, 1,
0, 2)\,.
\ed
The $T_{R_1}$ action on $A_T(0,1,0,2)$ differs from the $R_1$ action
by a sign in the action on the $(1/8,1/8)$ fields, i.e.\ the
twisted ground states of the $\Z_2$ orbifold on the $A_{\Z_2-{\rm orb}}(1/2,1/2,0,1)$
side. Therefore by comparison with~(\ref{d2h0+_z2r00})
\bd{d2h0-_z2tr00}
A_{D^{-}_2-{\rm orb}}\left(\frac 12, \frac 12, 0, 1\right) = 
A_{T_{R}-{\rm orb}}(0, 1, 0, 2)\,.
\ed
Comparison of~(\ref{d2h0+_z2r00}) with~(\ref{d2h0-_z2tr00}) also gives
a fairly natural explanation for the additional degree of freedom we
have due to discrete torsion.

If we mod out $\widetilde{\Z_2(R)}$ and the corresponding $D_2$ type
symmetries on $A_T(0,1,0,2)$, i.e.\ consider $\Z_2(R)$ on
$A_T(1/2,1/2,0,2)$ we only reproduce identities we have found already
above: $A_{D_2(T_R)-{\rm orb}}(0,1,0,1)=A_{R-{\rm orb}}(1/2,1/2,0,2)$
on~(\ref{z2rh0_d2tr}), as well as $A_{D_2(T_R^{(\prime)})-{\rm
orb}}(0,2,0,2)=A_{D_2^\pm-{\rm orb}}(1/2,1/2,0,2)$
on~(\ref{d200+_d2tr}) and~(\ref{d2h0-_d2trp}), respectively.

Next we discuss the action of $T_{R_1}$ on $A_T(0,1,0,1/2)$ instead of
$A_T(0,1,0,2)$.  In~(\ref{chch}) this exchanges the roles of $m_i$ and
$n_i$, such that compared to the action of $R_1$ on $A_T(0,1,0,2)$ we
now have additional signs on $(1/2,1/2)$ fields. In particular, only
one combination of $(1/2,1/2)$ fields is invariant, as well as three
of the twisted ground state combinations in $A_{\Z_2-{\rm
orb}}(0,1,0,1)$. We claim that this is the residual action of an
ordinary $\Z_4$ rotation on $A_T(0,1,0,1)$.  It acts by interchanging
the two circle factors of $A_T(0,1,0,1)$, but the generators of the
Hilbert space of the second factor are multiplied with an additional
sign. Indeed, this is exactly the $T_{R_1}$ action on a torus whose
lattice has an additional generator $(1/2,1/2)$ compared to $\Z^2$ for
$A_T(0,1,0,1)$, i.e.\ on $A_T(0,1,0,1/2)$.  Hence,
\bd{z4_z2tr00}
A_{\Z_4-{\rm orb}}(0, 1, 0, 1) = A_{T_R-{\rm orb}}\left(0, 1, 0, \frac 12\right).
\ed
Using~(\ref{z4_z2tr00}) we can further mod out $T_{R_2}$ on the
underlying torus theory of the above $A_{T_{R_1}-{\rm orb}}(0, 1, 0,
1/2)$.  This translates to a $\widetilde{\Z_2(R_2)}$ action on the
underlying torus theory of $A_{\Z_4-{\rm orb}}(0, 1, 0, 1)$, so
$$
A_{D_4^+-{\rm orb}}\left(0, 1, \frac 12, \frac 12\right) = 
A_{D_2(T_R)-{\rm orb}}\left(0, 1, 0, \frac 12\right).
$$
By~(\ref{d4h+_d4tr+}) we see that we have actually found a tricritical point:
\bt{d4tr+_d2tr}
A_{D_4^+-{\rm orb}}\left(0, 1, \frac 12,\frac 12\right) = 
A_{D_2(T_R)-{\rm orb}}\left(0, 1, 0, \frac 12\right)
= A_{D_4(T_R^\prime)^+-{\rm orb}}(0, 1, 0, 1)\,.
\et
We now rewrite~(\ref{z4_z2tr00}) as $A_{\Z_4-{\rm orb}}(0, 1, 0, 1) =
A_{T_{R_1}^\prime-{\rm orb}}(0, 1, 0, 1/2)$ and mod out by
$T_{R_2}^\prime$ on the underlying torus of the right hand side.
Analogously to the $T_{R_2}$ action on $A_{T_{R_1}-{\rm orb}}(0, t/2,
0, 2t)$ in~(\ref{z2tr00_d200-}), which induced a shift on the
underlying torus theory of $A_{D_2^--{\rm orb}}(0,t,0,t)$,
in~(\ref{z4_z2tr00}) we get a shift $T_{\delta^\prime},
\delta^\prime={1\over2}\left( {1\atop1} \right)$ on the underlying
torus theory of $A_{\Z_4-{\rm orb}}(0, 1, 0, 1)$. Then we obtain
\bd{z4_d2trp}
A_{\Z_4-{\rm orb}}(0, 1, 0, 2) = 
A_{D_2(T_R^\prime)-{\rm orb}}\left(0, 1, 0, \frac 12\right).
\ed
Back to the identification 
$A_{\Z_2-{\rm orb}}(0,1,0,1)=A_T(0,1,0,1/2)$ in ~(\ref{t_z2}) we now mod out
groups containing $\Z_4$ on the torus side. With the ordinary $\Z_4$
action we reproduce the above bicritical point~(\ref{z4_d2trp}), 
but in combination with $D_2(T_R^\prime)$, the $\Z_4$ generator
acts as a shift on the underlying torus theory of 
$A_{\Z_4-{\rm orb}}(0, 1, 0, 2)$ in~(\ref{z4_d2trp}):
\bds
A_{D_4(T_R^\prime)^+-{\rm orb}}\left(0, 1, 0, \frac 12\right)
&=& A_{\Z_4-{\rm orb}}(0, 1, 0, 4) \dlabel{d4trp+_z4}\\
A_{D_4(T_R^\prime)^--{\rm orb}}\left(0, 1, 0, \frac 12\right)
&=& A_{\Z_4-{\rm orb}}\left(0, 1, \frac 12, 1\right) \dlabel{d4trp-_z4}.
\eds
The latter identification is more easily understood when we mod out
symmetries on the tricritical point~(\ref{d200-_d2trp_z2}), as we will
do in section~\ref{t2}.

The effect of $D_4$ type actions is most easily understood from the
fact that by~(\ref{d200-_d2trpb}) the action of
$D_2(T_R^\prime)\subset D_4(T_R^\prime)^\pm$ on $A_T(0, 1, 0, 2)$ is
conjugate to that of $D_2^-\subset D_4^{-\pm}$.  Therefore,
\addtocounter{equation}{1}
\bea
A_{D_4(T_R^\prime)^+-{\rm orb}}(0, 1, 0, 2)&=& A_{D_4^{-+}-{\rm orb}}(0, 1, 0, 2) 
\label{d400-+_d4tr+}\\
 A_{D_4(T_R^\prime)^--{\rm orb}}(0, 1, 0, 2) &=&A_{D_4^{--}-{\rm orb}}(0, 1, 0, 2)\,.
\label{d400--_d4tr-}
\eea

\subsection{Series of multicritical points obtainable from (T1)}\label{t1}

From the multicritical points and lines determined so far we can find
further multicritical points by modding out further symmetries. By the
systematic procedure we followed above, this can only give something
new, if we use an identification obtained as intersection of
bicritical lines.  Moreover, because by the discussion at the
beginning of the section it suffices to use identifications containing
a toroidal theory, only (\ref{z2r00_d200+_t}) and~(\ref{t_d200-}) are
left to be discussed in this and the following section.

For the point~(\ref{z2r00_d200+_t}) only the identification
$A_T(0,1,0,4)=A_{D_2^+}(0,1,0,1)$ has not been used yet. By modding
out $\Z_2$ we yield~(\ref{d2h0+_d2tr_z2}) from~(\ref{z2r00_d200+_t}),
in particular $A_{\Z_2-{\rm orb}}(0,1,0,4)=A_{D_2^+-{\rm orb}}( 0,
1/2,1/2, 1/2)$.  Modding out a $\Z_2(R)$ action yields $A_{R-{\rm
orb}}(0,1,0,4)=A_{D_2^+-{\rm orb}}( 0, 2,0, 2)$
on~(\ref{z2r00_d200+}).  Note that this shows that $\Z_2$ and $R$ on
$A_T(0,1,0,4)$ both induce shifts on the underlying torus theory of
$A_{D_2^+}(0,1,0,1)$, namely $T_{\delta^\prime},
\delta^\prime={1\over2}\left( {1\atop1} \right)$, and $T_{\delta_1},
\delta_1={1\over2}\left( {1\atop0} \right)$, respectively.  The
combined action gives a trivial identity for $D_2^+$, and
$A_{D_2^--{\rm orb}}(0, 1, 0, 4)=A_{D_2^+-{\rm orb}}( 0, 1,1/2, 1)$ in
(\ref{d2h0+_d2tr_d200-}). Modding out $\Z_2(T_R), D_2(T_R)$ and
$D_2(T_R^\prime)$ reproduces the points at $t=2$ in
(\ref{z2tr00_d200-}),~(\ref{d200-_d2tr}), and~(\ref{d2h0-_d2trp}),
respectively. Modding out $\Z_4$ reproduces~(\ref{d4trp+_z4}). To
determine the result of modding out $D_4$ actions, note that by the
above the action of $R$ induces a shift $T_{\delta_1}$ on the
underlying torus theory of $A_{D_2^+}(0,1,0,1)$, so
from~(\ref{d4trp+_z4}) we obtain
\be\label{d4trp+_d4--}
A_{D_4^{--}-{\rm orb}}(0, 1, 0, 4)
= A_{D_4(T_R^\prime)^+-{\rm orb}}(0, 1, 0, 4)\,.
\ee
All the other choices of discrete torsion give trivial identities.
Modding out by $D_4(T_R^\prime)^\pm$ gives the same or a trivial identity
again.

Next we mod out $\widetilde{\Z_2(R)}$, i.e.\ $\Z_2(R)$ on
$A_T(1/2,1/2,0,4)$.  This interchanges the two circle factors of the
original $A_T(0,1,0,1)$ in $A_{D_2^+-{\rm orb}}(0, 1, 0, 1)$ above and
thus is equivalent to adding a $\Z_4$ generator to $D_2$. Therefore,
\bd{d40++_z2rh0}
A_{R-{\rm orb}}\left(\frac 12, \frac 12, 0, 4\right) =
A_{D_4^{++}-{\rm orb}}(0, 1, 0, 1)\,.
\ed
To mod out the corresponding $D_2$ actions we again use the above observation
that $\Z_2$ on $A_T(1/2,1/2,0,4)$ acts as $T_{\delta_1}$  on the 
underlying torus theory of $A_{D_2^+-{\rm orb}}(0, 1, 0, 1)$ to find
\bd{d2h0+_d400++}
A_{D_2^{+}-{\rm orb}}\left(\frac 12, \frac 12, 0, 4\right) = 
A_{D_4^{++}-{\rm orb}}(0, 1, 0, 2)\,,
\ed
and
$$
A_{D_2^--{\rm orb}}\left(\frac 12, \frac 12, 0, 4\right)
= A_{D_4(T_R^\prime)^+-{\rm orb}}(0, 1, 0, 2)\,,
$$
where the latter together with~(\ref{d400-+_d4tr+}) gives a tricritical
point
\bt{d4tr+_d20h-}
A_{D_2^--{\rm orb}}\left(\frac 12, \frac 12, 0, 4\right)
= A_{D_4(T_R^\prime)^+-{\rm orb}}(0, 1, 0, 2) = A_{D_4^{-+}-{\rm orb}}(0, 1, 0, 2)\,.
\et

\subsection{Series of multicritical points obtainable from (T2)}\label{t2}

We now discuss additional identifications that can be obtained from
(\ref{t_d200-}). The only identity not used up to now is
$A_T(0,1,1/2,1)=A_{D_2^--{\rm orb}}(0,1,0,1)$. If we mod out a $\Z_2$
action from the torus theory,~(\ref{t_d200-}) is transformed into
(\ref{d200-_d2trp_z2}), in particular we yield $A_{\Z_2-{\rm
orb}}(0,1,1/2,1)=A_{D_2^--{\rm orb}}(1/2,1/2,0,1/2)$. The $\Z_2$
action thus induces a shift $T_{\delta^\prime},
\delta^\prime={1\over2}\left( {1\atop1}\right)$ on the underlying
torus theory of $A_{D_2^--{\rm orb}}(0,1,0,1)$. The $R$ action on
$A_T(0,1,1/2,1)$ induces a shift as well, now by $T_{\delta_1},
\delta_1={1\over2}\left( {1\atop0}\right)$, yielding $A_{R-{\rm
orb}}(0,1,1/2,1)=A_{D_2^--{\rm orb}}(0,2,0,2)$
in~(\ref{z2tr00_d200-_d2tr_z2r}).  The combined $R$ and $\Z_2$ actions
thus yield a trivial identity for $D_2^-$ and $A_{D_2^+-{\rm
orb}}(0,1,1/2,1)=A_{D_2(T_R)-{\rm orb}}(0,1,0,4)$ on
(\ref{d200+_d2tr}). Modding out $\Z_4$ is equivalent to modding out
another $\Z_2$ action on $A_{\Z_2-{\rm orb}}(0,1,1/2,1) =A_{D_2^--{\rm
orb}}(1/2,1/2,0,1/2)$ which interchanges the circle factors of the
underlying geometric torus (i.e.\ $\Z_2$ invariant vertex operators
with $h=\bar h$). The action matches a $\widetilde{D_4}$ action on
$A_{D_2^--{\rm orb}}(1/2,1/2,0,1/2)$, where the additional $D_2^-$
invariant vertex operators as compared to $A_{D_2^--{\rm
orb}}(1/2,1/2,0,1)$ correspond to the $\Z_2$ twisted ground states of
$A_{\Z_2-{\rm orb}}(0,1,1/2,1)$. We thus obtain $A_{\Z_4-{\rm
orb}}(0,1,1/2,1)=A_{D_4(T_R^\prime)^--{\rm orb}}(0,1,0,1/2)$
reproducing~(\ref{d4trp-_z4}). Since by the above we know that $R_1$
on $A_T(0,1,1/2,1)$ induces a $T_{\delta_1}$ shift on the underlying
torus theory of $A_{D_2^--{\rm orb}}(0,1,0,1)$, it also follows that
\bd{d40h-_d4trp-}
A_{D_4^--{\rm orb}}\left(0,1,\frac 12,1\right)=
A_{D_4(T_R^\prime)^--{\rm orb}}(0,1,0,4)\,.
\ed
Flipping the sign of discrete torsion on both sides of the above
equivalence we find
$$
A_{D_4^+-{\rm orb}}\left(0,1,\frac 12,1\right)=
A_{D_4(T_R^\prime)^+-{\rm orb}}(0,1,0,4)\,,
$$
which together with~(\ref{d4trp+_d4--}) yields a tricritical point:
\bt{d4trp+_d4--_d40h+}
A_{D_4^+-{\rm orb}}\left(0,1,\frac 12,1\right)=A_{D_4(T_R^\prime)^+-{\rm orb}}(0,1,0,4)
= A_{D_4^{--}-{\rm orb}}(0, 1, 0, 4)\,.
\et
We now mod out $\widetilde{\Z_2(R)}$ on $A_T(0,1,1/2,1)$, i.e.\
$\Z_2(R)$ on $A_T(1/2,1/2,1/2,1)$. Similarly to~(\ref{d40++_z2rh0}) we find
\bd{d40--_d4tr-}
A_{R-{\rm orb}}\left(\frac 12, \frac 12, \frac 12, 1\right)=
A_{D_4^{--}-{\rm orb}}(0, 1, 0, 1) \,.
\ed
Because by the above, $\Z_2$ on $A_T(1/2,1/2,1/2,1)$ induces a shift
$T_{\delta^\prime}$ on the underlying torus theory of
$A_{D_4^{--}-{\rm orb}}(0, 1, 0, 1)$ in~(\ref{d40--_d4tr-}), we find
$$
A_{D_2-{\rm orb}}\left(\frac 12, \frac 12, \frac 12, 1\right)=
A_{D_4(T_R^\prime)^--{\rm orb}}(0, 1, 0, 2) \,.
$$
Together with~(\ref{d400--_d4tr-}) this gives another
tricritical point:
\bt{d4tr-_d2hh}
A_{D_2-{\rm orb}}\left(\frac 12, \frac 12, \frac 12, 1\right)
= A_{D_4(T_R^\prime)^--{\rm orb}}(0, 1, 0, 2) = A_{D_4^{--}-{\rm orb}}(0, 1, 0, 2)\,.
\et

\subsection{Multicritical points obtained from conjugate
$\Z_3, D_3, \Z_6$ and $D_6$ type actions}\label{z3mult}

We start by comparing all $\Z_3$ type symmetries of the $\SU(3)$ torus
theory at parameters $\tau=\rho=\omega$, $\omega:=e^{2\pi i/3}$.  The
generically conserved currents of the torus theory we call $j_1,j_2$,
and $k_1,k_2,k_3$ together with $l_\mu=k_\mu^\dagger, \mu\in\{1,2,3\}$
denote the additional vertex operators with dimensions $(h,\bar
h)=(1,0)$. The fields $j_\mu, k_\mu, l_\mu$ generate an $\SU(3)_1$
Kac--Moody algebra, and $\{k_\mu\}$, $\{l_\mu\}$ form closed orbits
under the ordinary $\Z_3$ action. In passing we remark that among all
possible $\Z_2$ symmetries of $A_T(\omega,\omega)$, those conjugate
only reproduce (\ref{z2rhh_z2}).

Among the $\Z_3$ actions on one hand we have the ordinary rotational
$\Z_3$ which leaves two fields $k_1+k_2+k_3$ and $l_1+l_2+l_3$
invariant, three fields $j^+=j_1+ij_2, k_1+\omega k_2+\omega^2 k_3,
l_1+\omega l_2+\omega^2 l_3$ have eigenvalue $\omega$. On the other
hand, the shift orbifold by $\delta={1\over2}(\lambda_1-\lambda_2)$
exhibits the same spectrum, where the $\lambda_i$ as usual denote a
basis of the lattice associated to the parameters $\tau=\rho=\omega$.
Here, $j_1,j_2$ are invariant, and $k_1,k_2,k_3$ have eigenvalue
$\omega$. We particularly see that the two $\Z_3$ actions are
conjugate, thus modding out $A_T(\omega,\omega)$ by these two
symmetries gives isomorphic theories.  The shift orbifold again
produces a torus theory with same parameter $\tau=\omega$, but $\rho$
reduced by a factor of three; in the following we use
$\alpha:=1/2+i3\sqrt3/2$ which is related to $\omega/3$ by the M\o
bius transformation $\mathbb{T}^2S$ and state
\bd{t_z3}
A_{\Z_3-{\rm orb}} \left(\frac 12,\frac{\sqrt{3}}2, \frac 12, \frac{\sqrt{3}}2\right) =
A_T\left(\frac 12, \frac{\sqrt{3}}2, \frac 12, \frac{3\sqrt{3}}2\right) .
\ed
We will now mod out additional symmetries on both sides of the above
equality. Only those of order two give new identifications.  Note that
both $R_2$ and the ordinary $\Z_2$ on $A_T(\omega,\omega)$ interchange
the two $\Z_3$--invariant $(1,0)$ fields $k_1+k_2+k_3$ and
$l_1+l_2+l_3$.  Thus $R_2,\Z_2$ must act as $R_1,R_2$ on the torus
theory $A_T(\omega,\alpha)$. Study the action on the charge lattice to
check that the order above is indeed correct. This means that the
$R_1$ action on $A_T(\omega,\omega)$ must induce the ordinary $\Z_2$
action on $A_T(\omega,\alpha)$. In particular, the fields
$k_1+k_2+k_3$ and $l_1+l_2+l_3$ are multiplied by $-1$ under
$R_1$. Here we can confirm our result of the discussion of lattice 7:
The signs obtained there occur in a completely natural way in the
present example.

All in all for the $\Z_2$ actions on $A_T(\omega,\omega)$ compared to
$A_T(\omega,\alpha)$ we have found $(R_1,R_2,\Z_2)\mapsto
(R_2,\Z_2,R_1)$ and therefore directly obtain the following bicritical
points:
\bds
A_{D_3(R_1)-{\rm orb}}\left(\frac 12, \frac{\sqrt{3}}2, \frac 12, \frac{\sqrt{3}}2\right) 
& = & A_{\Z_2-{\rm orb}}\left(\frac 12, \frac{\sqrt{3}}2, \frac 12,
\frac{3\sqrt{3}}2\right),
\dlabel{d3h_z2}\\
A_{D_3(R_2)-{\rm orb}}\left(\frac 12, \frac{\sqrt{3}}2, \frac 12, \frac{\sqrt{3}}2\right) 
& = & A_{R_1-{\rm orb}}\left(\frac 12, \frac{\sqrt{3}}2, \frac 12,
\frac{3\sqrt{3}}2\right),
\dlabel{d3h_z2rhh}\\
A_{\Z_6-{\rm orb}}\left(\frac 12, \frac{\sqrt{3}}2, \frac 12, \frac{\sqrt{3}}2\right) 
& = & A_{R_2-{\rm orb}}\left(\frac 12, \frac{\sqrt{3}}2, \frac 12,
\frac{3\sqrt{3}}2\right),
\dlabel{z6_z2rhh}\\
A_{D_6-{\rm orb}}\left(\frac 12, \frac{\sqrt{3}}2, \frac 12, \frac{\sqrt{3}}2\right) 
& = & A_{D_2-{\rm orb}}\left(\frac 12, \frac{\sqrt{3}}2, \frac 12,
\frac{3\sqrt{3}}2\right).
\dlabel{d6h+-_d2hh}
\eds

\section{Product theories within the moduli space}\label{product} 

If our description of nonisolated components of $\mathcal{C}^2$ is
complete, it must be possible to find all nonisolated components known
so far. In particular, we should consider tensor products of known
models.  The simplest case is the product of two models with central
charge $c=1$. The possible factor theories then are $A^{c=1}(r),
A^{c=1}_{\rm orb}(r), A^{c=1}_T, A^{c=1}_O$, and $A^{c=1}_I$,
corresponding to compactification on a circle with radius $r$, its
$\Z_2$ orbifold or one of the three isolated components of the $c=1$
moduli space, respectively.  Models containing one of the latter three
factor theories are exceptional but of course easily constructed, as
was mentioned in section~\ref{tori}.  Moreover,
\begin{eqnarray*}
A^{c=1}(r)\otimes A^{c=1}(r^\prime)
&=& A_T\left(0,{ {r^\prime\over r}}, 0, rr^\prime\right),
\\
A^{c=1}(r)\otimes A^{c=1}_{\rm orb}(r^\prime)
&=& A_{R_1-{\rm orb}}\left(0,{ {r^\prime\over r}}, 0,
rr^\prime\right),
\end{eqnarray*}
and
$$
A^{c=1}_{\rm orb}(r)\otimes A^{c=1}_{\rm orb}(r^\prime)
= A_{D_2^+-{\rm orb}}\left(0,{ {r^\prime\over r}}, 0, rr^\prime\right)
$$
are obvious (see~(\ref{circle2}),~(\ref{lattice6}),~(\ref{lattice80})).

Using the results of~\cite{dix}, nonisolated components of the moduli
space can also be obtained by tensoring $N=1$ superconformal field
theories $A_{\bullet}^{c=3/2}(r)$ with central charge $c=3/2$ with the
unique unitary conformal field theory at $c=1/2$. In this section we
discuss how the resulting models $A_{\bullet}^{M}(r)$ can be found
within the components of $\mathcal{C}^2$ we have determined in
section~\ref{part}.

By~\cite{dix}, the moduli space of $N=1$ superconformal field theories
with $c=3/2$ contains five connected lines. The \emph{circle line}
$A_{\rm circ}^{c=3/2}(r)$ is obtained from the $c=1$ circle theories by
adding one Majorana fermion, i.e.\ tensoring with the unique unitary
conformal field theory at $c=1/2$, the Ising model.  Since the tensor
product of two Ising models has a bosonic description as $\Z_2$
orbifold of the $c=1$ circle theory at radius $r^\prime=\sqrt{2}$, by
the discussion of lattice 6 we directly obtain
$$
A_{\rm circ}^{M}(\sqrt2 r) = A_{\rm orb}^{c=1}(\sqrt2)\otimes A^{c=1}(\sqrt2 r)
= A_{R_2-{\rm orb}}(0,r,0,2r)\,.
$$
The other four lines in the $c=3/2$ moduli space are obtained as
orbifold models of $A_{\rm circ}^{c=3/2}(r)$.  The ordinary $\Z_2$
orbifold generates the so-called \emph{orbifold line} $A_{\rm
orb}^{c=3/2}(r)$.  For the fermions the orbifold procedure effectively
only exchanges boundary conditions, which we forget about in our $c=2$
purely bosonic language. Therefore, we can regard $\Z_2$ as only
acting on the second circle factor of $A_{\rm circ}^{M}(\sqrt2
r)=A_{R_2-{\rm orb}}(0,r,0,2r)$.  This amounts to modding out an $R_1$
action, i.e.\
$$
A_{\rm orb}^{M}(\sqrt2 r)=A_{D_2^+-{\rm orb}}(0,r,0,2r)\,.
$$
Note that by the results of section~\ref{meet} and in agreement
with~\cite{dix} the only intersection point of the above lines is
situated on~(\ref{z2r00_d200+}):
$$
A_{\rm circ}^{M}(2)=A_{\rm orb}^{M}(1)\,.
$$
The \emph{superaffine line} $A_{s-a}^{c=3/2}(r)$ is the orbifold of
$A_{\rm circ}^{c=3/2}(r)$ by the $\Z_2$ type group generated by
$S_\delta:=t_\delta\,(-1)^{F_S}$. Here, $t_\delta=e^{2\pi
ip{\delta\over\sqrt2}}$ is the shift orbifold on the bosonic $c=1$
theory, and $(-1)^{F_S}$ is the spacetime fermion number
operator. $(-1)^{F_S}$ acts by multiplication with $-1$ on the Ramond
sector and trivially on the Neveu--Schwarz sector of the theory.  To
determine $A_{s-a}^{M}(\sqrt2 r)$, we trivially continue the action of
$S_\delta$ to $A_{\rm circ}^{M}(\sqrt2 r)$.  Then $S_\delta$ remains to
act as ordinary shift orbifold on the second factor theory in
$A_{\rm circ}^{M}(\sqrt2 r)$, the $c=1$ circle theory at radius $\sqrt2
r$.  On the first factor, we have the action of $(-1)^{F_S}$ on one of
the Majorana fermions.  We use the bosonic description as $\Z_2$
orbifold of the $c=1$ circle theory at radius $\sqrt2$.  Here, the
Ramond sector is built on those Hilbert space ground states with odd
label of the momentum mode. Thus on the underlying $c=1$ circle
theory, $(-1)^{F_S}$ acts as shift orbifold as well.  This means that
$A_{s-a}^{M}(\sqrt2 r)$ can be obtained as shift orbifold by
$T_{\delta^\prime}$, $\delta^\prime={1\over\sqrt2}\left( {1\atop
r}\right)$ on the underlying torus theory $A_T(0,r,0,2r)$ of
$A_{\rm circ}^{M}(\sqrt2 r)$:
$$
A_{s-a}^{M}(\sqrt2 r) = A_{R_2-{\rm orb}} 
\left(\frac 12, \frac r2, 0, r\right).
$$
The \emph{superorbifold line} $A_{s-{\rm orb}}^{c=3/2}(r)$ is a $D_2$
type orbifold of $A_{\rm circ}^{c=3/2}(r)$ by the group generated by
the ordinary $\Z_2$ action and $S_\delta$. Since by the above $\Z_2$
and $S_\delta$ act as reflection $R_1$ and shift $T_{\delta^\prime}$
on the underlying torus theory $A_T(0,r,0,2r)$ of $A_{\rm
circ}^{M}(\sqrt2 r)$, respectively, we find
$$
A_{s-{\rm orb}}^{M}(\sqrt2 r) = A_{D_2^+-{\rm orb}} 
\left(\frac 12, \frac r2, 0, r\right).
$$
By the results of section~\ref{meet} we see that only the
superorbifold line intersects one of the other three lines discussed
so far, namely in~(\ref{d2h0+_z2r00}):
$$
A_{s-\rm orb}^{M}(\sqrt2)=A_{\rm circ}^{M}(\sqrt2)\,.
$$
This agrees with the results of~\cite{dix}.  Finally, the
\emph{orbifold-prime line} $A_{\rm orb^\prime}^{c=3/2}(r)$ is obtained by
modding out $S_R:=(-1)^{F_S}\cdot(-1)$ from $A_{\rm circ}^{c=3/2}(r)$,
where $(-1)$ is the generator of the ordinary $\Z_2$ action.  In
particular, for the partition functions of orbifold and
orbifold--prime theories, one has the relation
\be\label{ellgen} 
Z_{{\rm orb}^\prime}^{c=3/2}(r) = Z_{\rm orb}^{c=3/2}(r) -3\,.
\ee
Since the generator of the ordinary $\Z_2$ action on $A_{\rm
circ}^{M}(\sqrt2 r)$ acts as reflection $R_1$, and $(-1)^{F_S}$ is the
shift orbifold on the underlying $c=1$ circle theory at radius
$\sqrt2$ of the first factor in $A_{\rm circ}^{M}(\sqrt2)$, $S_R$ acts
as $T_{R_1}$ on the underlying torus theory $A_T(0,r,0,2r)$ of $A_{\rm
circ}^{M}(\sqrt2 r)$. Therefore,
$$
A_{\rm orb^\prime}^{M}(\sqrt2 r)
= A_{D_2(T_R)-{\rm orb}}(0,r,0,2r)\,.
$$
Concerning intersections of the orbifold--prime line with the other
lines discussed above, again we are in exact agreement with the
results of~\cite{dix}: we find multicritical points
on~(\ref{d200+_d2tr}) and (\ref{z2rh0_d2tr}), namely
$$
A_{\rm orb^\prime}^{M}(2)=A_{s-{\rm orb}}^{M}(2)\,, \qquad
A_{\rm orb^\prime}^{M}(1)=A_{s-a}^{M}(2)\,.
$$
It is a straightforward calculation to check~(\ref{ellgen}) for our
$c=2$ models, i.e.\
$$
Z_{D_2(T_R)-{\rm orb}}(0,r,0,2r) = Z_{D_2^+-{\rm orb}}(0,r,0,2r) -
3Z_{\rm Ising}
$$
from~(\ref{lattice80}),~(\ref{lattice11}), and~(\ref{Ising}).

The above in particular gives a geometric interpretation in terms of
crystallographic orbifolds to all the nonisolated orbifolds discussed
in~\cite{dix}.

\section{Conclusions}

We have explicitly constructed the parameter spaces and the one loop
partition functions of the sixteen types of crystallographic orbifold
conformal field theories of toroidal theories with central charge
$c=2$. Taking into acount all possible choices of the B--field and all
values of discrete torsion, this yields $28$ different components of
the moduli space $\mathcal{C}^2$ of unitary conformal field theories
with central charge $c=2$.  We have argued that this way, apart from
the exceptional cases related to the binary tetrahedral, octahedral
and icosahedral subgroups of $\SU(2)$, we get all the nonisolated
irreducible components of the moduli space that can be obtained by an
orbifold procedure.  In the construction of the various theories some
unexpected effects of the B--field have occured which might lead to a
better understanding of its properties, also for higher dimensional
cases.

We have determined all the multicritical points and lines of the $28$
components of $\mathcal{C}^2$ constructed before. We have found \crl
bicritical lines and \crit multicritical points, among them \qcr
quadrucritical and \tcr tricritical points.  We have proven
multicriticality on the level of the operator algebra for all these
lines and points.  The case by case study also sheds some light on the
effect of discrete torsion.

Drawing a picture of the moduli space $\mathcal{C}^2$ one will notice
a complicated graph like structure with a lot of loops. In particular,
by our analysis of multicritical points, all but four of the
irreducible crystallographic components of the moduli space are
directly or indirectly connected to the moduli space of toroidal
theories.  The remaining four components are
$\mathcal{C}_{D_4^{+-}-{\rm orb}}^{(0)}, \mathcal{C}_{D_6^{\pm}-{\rm
orb}}^{(0)}, \mathcal{C}_{D_3(R)-{\rm orb}}^{(0)}$.

We have related our results to those on $c=3/2$ superconformal field
theories~\cite{dix}. This was done by determining the tensor products
of the five continuous lines of $c=3/2$ superconformal field theories
discussed in~\cite{dix} with an Ising model in terms of our
description of $\mathcal{C}^2$. All multicritical points in the
$c=3/2$ moduli space are reidentified by our results on
$\mathcal{C}^2$. In particular, this gives geometric interpretations
to all nonisolated orbifolds discussed in~\cite{dix} in terms of
crystallographic orbifolds.

A discussion of the exceptional components of $\mathcal{C}^2$ is not
carried out in this work. By our results, these would yield the only
possible examples of asymmetric orbifold conformal field
theories~\cite{nsv87} with $c=2$ and therefore should be studied
separately.  Neither do we touch the determination of isolated
components of the moduli space, which is expected to be even more
involved.  Apart from that, our results do not give a complete
classification of unitary conformal field theories with central charge
$c=2$, since we are lacking a theorem which would tell us that all
nonisolated components of the moduli space may be obtained by some
orbifold procedure from a subspace of the toroidal component.  It
would also be interesting to determine those theories in
$\mathcal{C}^2$ which admit supersymmetry.

\acknowledgments

It is a pleasure to thank Werner Nahm, not only for driving our
interest to the problems discussed in this paper. Without the
countless discussions with him this work would not have been
possible.

S.D. would like to thank M.~Soika for many helpful
discussions.

S.D. was supported by the DAAD. Part of the work was
also supported by TMR.

\end{document}